\documentclass[a4paper,11pt]{article}
\include{definitions}
\pdfoutput=1 

\usepackage{jheppub} 

\usepackage[T1]{fontenc} 
\usepackage{amssymb,amsmath,amsfonts}
\usepackage{mathtools}
\usepackage{mathrsfs}
\usepackage{bbm}
\usepackage{slashed}
\usepackage{nicefrac}
\usepackage{graphicx}
\usepackage{caption}
\usepackage{subcaption}
\usepackage{esint}
\usepackage{verbatim}

\usepackage{graphicx}
\usepackage[dvipsnames]{xcolor}
\usepackage{array}

\newcommand{\eqn}[1]{Eq.~\eqref{#1}}

\long\def\comment#1{ }

\def\k{{\boldsymbol k}}

\def\q{{\boldsymbol q}}
\def\p{{\boldsymbol p}}

\def\r{{\boldsymbol r}}
\def\A{{\boldsymbol A}}
\def\x{{\boldsymbol x}}

\def\0{{\boldsymbol 0}}

\def\k{{\boldsymbol k}}

\def\n{{\boldsymbol n}}
\def\x{{\boldsymbol x}}

\def\p{{\boldsymbol p}}

\def\r{{\boldsymbol r}}

\def\A{{\boldsymbol A}}
\def\C{{\boldsymbol C}}
\def\Sc{\mathcal{S}}
\def\Cc{\mathcal{C}}
\def\cC{\mathcal{C}}

\title{Wilson line correlators beyond the large-$N_c$}

\author[a]{Johannes Hamre Isaksen}
\author[a]{and Konrad Tywoniuk}
\affiliation[a]{Department of Physics and Technology, University of Bergen, 5007 Bergen, Norway}

\emailAdd{johannes.isaksen@uib.no}
\emailAdd{konrad.tywoniuk@uib.no}

\abstract{
We study hard $1\to 2$ final-state parton splittings in the medium, and put special emphasis on calculating the Wilson line correlators that appear in these calculations. As partons go through the medium their color continuously rotates, an effect that is encapsulated in a Wilson line along their trajectory. When calculating observables, one typically has to calculate traces of two or more medium-averaged Wilson lines. These are usually dealt with in the literature by invoking the large-$N_c$ limit, but exact calculations have been lacking in many cases. In our work, we show how correlators of multiple Wilson lines appear, and develop a method to calculate them numerically to all orders in $N_c$. Initially, we focus on the trace of four Wilson lines, which we develop a differential equation for. We will then generalize this calculation to a product of an arbitrary number of Wilson lines, and show how to do the exact calculation numerically, and even analytically in the large-$N_c$ limit. Color sub-leading corrections, that are suppressed with a factor $N_c^{-2}$ relative to the leading scaling, are calculated explicitly for the four-point correlator and we discuss how to extend this method to the general case. These results are relevant for high-$p_T$ jet processes and initial stage physics at the LHC.
}

\begin{document}
\maketitle
\section{Introduction}
\label{sec:intro}

One of the primary reasons for colliding heavy ions with ultra-relativistic energies is to probe QCD matter in an extremely hot and dense phase, called the quark-gluon plasma (QGP).
There are multiple ways to probe and learn about the properties of QGP, utilizing the properties of bulk particle production and rare probes. In one example of the latter category, the heavy-ion collision involves a hard partonic sub-collision that produces hard partons that propagate through the medium and escape to the detectors as jets. The study of how the properties of these jets change as they go through the medium, colloquially referred to as ``jet quenching,'' is a versatile tool to study hot QCD matter \cite{dEnterria:2009xfs,Majumder:2010qh,Mehtar-Tani:2013pia}.

Experiments at RHIC \cite{Adams:2005dq,Adcox:2004mh} and the LHC \cite{Aamodt:2010jd,Khachatryan:2016odn,Chatrchyan:2011sx,Aad:2010bu,Abelev:2013kqa} colliders have found strong suppression of high-$p_T$ particles in heavy-ion collisions compared to proton-proton collisions, which is interpreted as a clear sign of the energy loss of jets that suffer final-state interactions with the surrounding QGP. On the theoretical side, this is interpreted in terms of radiative energy loss, where particles in the jet lose energy through medium-induced emission of gluons that end up outside of the reconstructed jet cone, and elastic drag. For large media, as typically encountered in central to semi-central lead-lead collisions, it is the former process that dominates the total lost energy. 

The energy loss process for single partons is well understood since many years, see, e.g., \cite{Baier:1994bd,Baier:1996sk,Baier:1996kr,Baier:1998yf,Zakharov:1996fv,Zakharov:1997uu,Wiedemann:2000za}. However, a jet is a more complicated composite object consisting of several hard partons. A hard parton propagating through the medium will typically undergo several splittings, resulting in a multi-parton state that will interact differently with the medium compared to how the individual partons would. Such splittings can occur as long as the scale of the splittings, for instance the generated relative transverse momentum in the splitting, is bigger than what the medium can supply through multiple scattering. In particular, the modifications of effects of color coherence play an important role in determining which emissions will be resolved by the medium and contribute toward the total energy loss \cite{Mehtar_Tani_2018,Mehtar-Tani:2019tvy,Caucal:2018dla,Caucal:2020xad}. Instead of focusing on single partons, we will study a hard parton splitting into two, and their subsequent propagation through the medium. This is certainly a better approximation of a real jet than a single parton, and has the additional advantage that one can build up jets from several partons by consecutive $1\to2$ splittings. 

Previous studies of such processes focused mostly on a hard photon splitting into a quark-antiquark pair \cite{Mehtar_Tani_2018,Dom_nguez_2020} and invoked the large-$N_c$ approximation to obtain analytical formulas. In this work, we consider three generic QCD splitting processes that involve up to eight correlated Wilson lines in the fundamental representation, in the case of gluon splitting into two daughter gluons. 
Our specific improvement concerns a more precise way to calculate correlators of Wilson lines that often appear in these calculations, and it can, in principle, be extended for an arbitrary number of propagating particles through the medium.

To give a general flavor of how our procedure works, recall that a matrix element generally involves several propagators that resum multiple scattering through Wilson lines $V$, which extend along the trajectories in the medium. Ignoring some factors irrelevant for the present discussion, the matrix element squared will take the following simplified form
\begin{align}
\label{eq:generic-structure-amplitude-squared}
\left\langle |\Mc|^2 \right\rangle &\sim \langle\tr[V^\dagger V \ldots V^\dagger V]\ldots \tr[V^\dagger V \ldots V^\dagger V]\rangle\,,
\end{align}
where the angular brackets denote an average over medium configurations. For a generic $1\to 2$ process, the amplitude squared can be reduced to a product of two-, three- and four-point correlators \cite{Blaizot_2013,Apolinario:2014csa}. 
To calculate these processes it is imperative to know the form of the Wilson line correlator appearing on the right hand side of \eqref{eq:generic-structure-amplitude-squared},
which we will denote by the letter $C^K$ for correlator, where the subscript $K$ refers to the number of traces. If you assume that the number of colors $N_c$ is large the calculation of these correlators usually simplifies sufficiently to be possible to calculate. Namely, the leading $N_c$ scaling emerges from simplifying the medium averages to $\langle \tr[V^\dag V\ldots V^\dag V]\rangle \ldots \langle \tr[V^\dagger V \dots V^\dagger V]\rangle$, which scales like $N_c^K$. However, since $N_c=3$ is not a very large number it is sensible to ask whether this approximation is sound or not. As we will see, the terms that are discarded by performing the large-$N_c$ approximation will be smaller than the other terms by a factor $\sim 1/N_c^2 \simeq 10 \%$ for typical situations. However, evaluating the correlators at large times, could lead to big discrepancies between the finite and large-$N_c$ calculations.

In this paper we will develop a method for calculating correlators of an arbitrary number of Wilson lines at finite $N_c$, which casts their evolution and mixing in terms of a coupled evolution equation in time (referring to their trajectories through the medium). This reduces the complexity of the formulation compared to previous calculations of multi-Wilson line correlators, see \cite{Kovner_2001} for a technique based on diagonalization of the evolution matrix and  \cite{Dominguez:2011wm,Apolinario:2014csa} for an iterative procedure. The derivation of the evolution matrix culminates in Eq.~\eqref{master-diff-eq}. This allows us to evaluate these correlators at an arbitrary time, and can be addressed using numerical techniques. We also consider in detail the large-$N_c$ approximation, which leads to a striking simplification of the dynamics since all higher-order correlators can be calculated using two-point correlators (dipoles) and their convolutions. Furthermore, we have computed the sub-leading correction in color. Considering again the generic correlator $C^K$ from Eq.~\eqref{eq:generic-structure-amplitude-squared} above, the generic expansion in $N_c$ takes the following form,
\begin{equation}\label{eq:generic-largeNc-expansion}
    C^K = N_c^K \hat{C}^{K}_{\text{leading $N_c$}}  + N_c^{K-2} \hat{C}^{K}_{\text{sub-leading $N_c$}} + \mathcal{O}(N_c^{K-4}) \,,
\end{equation}
where the two first terms can be found analytically (the hat over the correlators imply that we have explicitly extracted their leading $N_c$ behavior). It turns out that, in many cases, $C^K_\text{sub-leading $N_c$}$ is essential to recover the correct long-time behavior of the correlators.

We will explore how big the error is by comparing the exact results to the large-$N_c$ approximation in realistic settings in high-energy jet splittings. We mainly consider hard emissions early in the medium, i.e. at scales much larger than those provided by the medium, and therefore we neglect any broadening of the particles. The daughters are traversing the medium at a fixed angle, or ``tilt'', given by the kinematics of the hard splitting (we fix our coordinate system so that the parent particle has zero angle). For splittings where at least one of the daughters becomes very soft or is being emitted at a large angle, one should also allow for additional transverse momentum broadening, as done in \cite{Blaizot_2013,Apolinario:2014csa}, albeit only in the large-$N_c$ approximation. We have left this additional complication for future work.

Our calculation is also very pertinent for improving our understanding of color dynamics in the medium, for instance in the context of multi-gluon emissions with overlapping formation times \cite{Arnold:2019qqc} and to understand hadronization after exiting the QGP \cite{Zakharov:2018hfz}.
In the process of evaluation of the multi-Wilson line correlators, the only assumption made is the exact form of the medium average, see Eq.~\eqref{eq:med-avg}, which is also employed in other contexts than for a thermal medium, see, e.g., \cite{Hatta:2020wre} for calculating such correlators on the lattice. Therefore, although we have derived our method of calculating Wilson line correlators in the context of jet quenching, it is a general result that can be applied in more branches of QCD. One concrete example refer to initial state physics, where multi-particle production is considered an important channel to verify saturation effects in the nuclei \cite{Kovner_2001,Jalilian-Marian:2004vhw,Iancu:2011ns}. Furthermore, sub-leading corrections in color have also been considered in the context of high-energy QCD evolution at next-to-leading order \cite{Lappi:2020srm}. Finally, the generic color structure of high-energy QCD events is actively studied \cite{Dominguez:2011wm,Dominguez:2012ad}. It is also interesting to note that sub-leading color corrections have been considered in the context of improving parton showers in the vacuum, see, e.g., \cite{Nagy:2012bt,Hamilton:2020rcu}.

Let us briefly outline the structure of the paper.
Section \ref{sec:basic-elements} introduces the notation and formalism we will make use of throughout the paper. In Sec.~\ref{sec:split-processes} we will consider three examples of splitting processes that lead to Wilson line correlators: a photon producing a quark-antiquark pair, a quark emitting a gluon and a gluon splitting into two gluons. Those processes will provide the motivation for the rest of the calculation in the paper. In Sec.~\ref{sec:four-lines}, we will go into detail about calculating the simplest of the Wilson lines structures from Sec.~\ref{sec:split-processes}, which is a trace of four lines.
Here, we also develop a method to compute the color sub-leading corrections, corresponding to the second term on the right hand side in Eq.~\eqref{eq:generic-largeNc-expansion}.
Thereafter, in Sec.~\ref{sec:Wilson} we will generalize the method used in Sec.~\ref{sec:four-lines} to correlators of an arbitrary number of Wilson lines, and show how one can always make a system of differential equations to describe these structures. This section contains the main theoretical results of the paper. The formulas developed in Sec.~\ref{sec:Wilson} are used to calculate the more complicated Wilson line structures appearing in Sec.~\ref{sec:split-processes}. We will show how the calculations simplify in the large-$N_c$ approximation, and use numerical evaluation to compare the approximate results to the exact ones.

\section{Basic elements and notation}
\label{sec:basic-elements}

We will assume that the partons propagating through the medium are highly energetic and  travelling on the light-cone almost strictly in the positive $z$ direction. In light-cone (LC) coordinates it will have momentum $(p^+,p^-,\p)$, where $p^+=(p^0+p^3)/2$ is identified with the LC energy $E\equiv p^+$, $p^- = p^0 - p^3$ is negligible and $\p$ is the transverse momentum. The parton interacts with the medium, which is modelled by a classical background gauge field $A^{\mu,a}(t,\r)$. The interaction of the parton with the classical field leads to transverse momentum broadening and energy loss. 
The interactions can be resummed using a framework developed by Baier-Dokshitzer-Mueller-Peigné-Schiff \cite{Baier:1994bd,Baier:1996sk,Baier:1996kr,Baier:1998yf} and Zakharov \cite{Zakharov:1996fv,Zakharov:1997uu}, and is known as the BDMPS-Z formalism. For small media, where interactions are rare, this is equivalent with considering only one interaction, known as the Gyulassy-Levai-Vitev (GLV) \cite{Gyulassy:2000er} approximation.

It is possible to construct Feynman rules from the BDMPS-Z approach, with special in-medium propagators and vertices \cite{Mehtar_Tani_2018}. In this formulation a highly energetic parton travelling through the medium can be described by the propagator
\beq
\label{eq:prop-G}
(\x|\Gc_R(t,t_0)|\x_0)=\Theta(t-t_0)\, \int^{\x}_{\x_0} \cD\r \exp\left[i\frac{E}{2}\int_{t_0}^t \dd s\,  \dot \r^2(s) \right]\, V_R(t,t_0;\r(t))\,.
\eeq
In this expression $V_R$ is a Wilson line in the representation $R$, which is given by
\begin{equation}\label{Wilson-definition}
V_R\left(t, t_{0} ;\mathbf r (t)\right)=\mathcal{P} \exp \left[i g \int_{t_{0}}^{t} \mathrm{d} s \, A^a(s, \mathbf{r}(s))T^{a}_R\right] \,,
\end{equation}
where the symbol $\mathcal{P}$ enforces path ordering.
A quark transforms in the fundamental representation, so the group generator is $T_F^a\equiv t^a_{ij}$. Similarly, a gluon transforms in the adjoint representation, and its group generator is $T_A^a\equiv (T^a)^{bc}=-if^{abc}$. The final results in this paper will mainly concern fundamental lines, which we will denote by $V \equiv V_F$. Similarly, we will write the adjoint lines as $U \equiv V_A$. Focusing on fundamental lines is sufficient, since one can always transform adjoint Wilson lines to fundamental ones through the identity
\begin{equation} 
\label{eq:adj-fund-Wlines}
U^{a b}= 2 \operatorname{tr}\left[ t^{a} V t^{b} V^{ \dagger}\right]=U^{\dagger ba}\,.
\end{equation}
In the absence of interactions, i.e. when the Wilson line is evaluated at $g = 0$, we simply get
\begin{equation}
\label{eq:G0}
(\x | \Gc_0(t,t_0) |\x_0) \equiv \Gc_0(\x-\x_0,t-t_0)= \Theta(t-t_0)\frac{E}{2\pi i(t-t_0)} \rme^{i \frac{E}{2} \frac{(\x-\x_0)^2}{(t-t_0)}} \,,
\end{equation}
which is a representation of the retarded part of the Feynman propagator ($E>0$).

As mentioned in the introduction, the matrix element describing final-state interactions in the QGP will involve one or more propagators of the form in Eq.~\eqref{eq:prop-G}. Hence, on the level of the matrix element squared, we have to compute correlators of such lines averaged over all possible medium configurations.
The medium average is indicated by $\langle \dots \rangle$, and we assume that the correlator of the medium fields takes the form
\begin{equation}\label{eq:med-avg}
\langle A^{a}(t, \r) A^{b}(t', \r')\rangle = \delta^{a b} n(t) \delta (t-t') \gamma\left(\boldsymbol{r}-\boldsymbol{r}^{\prime}\right) \,,
\end{equation}
which corresponds to the Gaussian noise approximation. Here, $n(t)$ is the (time-dependent) density of scattering centers in the medium and
\begin{equation}
    \gamma(\r) = \int \frac{\rmd^2 \q}{(2\pi)^2} \, \rme^{ i \q \cdot \r} \frac{\rmd^2 \sigma_{\rm el}}{\rmd^2 \q} \sim g^2 \int\frac{\rmd^2 \q}{(2\pi)^2} \,  \frac{\rme^{ i \q \cdot \r}}{\q^4} \,,
\end{equation} 
is the Fourier transform of the in-medium elastic scattering potential, where the infrared behavior of the potential is regulated by an in-medium screening mass.
The delta function in time indicates that we have assumed the medium interactions to be instantaneous. In many cases it will be convenient to define
\begin{equation} 
\label{eq:sigma-gamma}
\sigma(\r) = g^2 \big[\gamma(\bm 0) - \gamma(\bm r)\big] \,.
\end{equation}
The form of the function $\sigma$ depends on how the medium is modelled. The two main ways of calculating this is through the Gyulassy-Wang model \cite{Wang:1991xy} or through Hard Thermal Loop theory \cite{Aurenche:2002pd}. These models differ mainly in how infrared screening is implemented when $q_\perp \to 0$. In this paper, we will however work in the harmonic oscillator approximation, which accounts for multiple soft interactions. In this case, the potential $\sigma(\r)$ can be cast as
\begin{align} \label{eq:harmonic-pot}
C_R n \sigma(\bm r) 
\simeq \frac{1}{4} \r^2 \hat q_R(t)\,, 
\end{align}
where 
\begin{equation} 
\label{eq:qhat-def}
\hat q_R = C_R n g^2 \int^{q^{\rm max}_{\perp}} \frac{\rmd^2 \q}{(2\pi)^2} \, \q^2 \frac{\rmd^2 \sigma_{\rm el}}{\rmd^2 \q} \,,
\end{equation}
is the jet quenching coefficient where $R$ denotes the color representation of the Wilson lines.  For the fundamental and adjoint representations we have $C_F=\frac{N_c^2-1}{2N_c}$ and $C_A=N_c$, respectively. In this paper we will use $\qhat = \qhat_F$ unless otherwise stated. In \eqn{eq:qhat-def} we have explicitly introduced a UV cut-off to regularize the integral. A more systematic approach to the regularization of the integral, and the extension beyond the soft scattering approximation, has been pursued in Refs.~\cite{Mehtar-Tani:2019tvy,Mehtar_Tani_2020,Barata:2020rdn}.

We stress that the approximation in \eqref{eq:harmonic-pot} is not necessary in order to solve numerically the system of equations for arbitrary $n$-point correlators, but it is very useful to employ to compare these exact results to analytical calculations of the leading and sub-leading color correlators.

In the current work, we will focus on hard $1\to 2$ splitting processes in the medium, where the initial particle has energy $E$ and the two splitting products carry, respectively, $\omega_1=(1-z)E$ and $\omega_2=zE$. This is formally equivalent to setting the energy of the mother particle, $E \to \infty$, and considering a finite momentum sharing fraction $ 0 \ll z \ll 1$. 
These conditions enforce that both the mother and daughter particles travel on classical paths. Concretely, the trajectory of a particle in the medium between time $t_0$ and $t$,  given by the the propagator $(\x|{\Gc}(t,t_0) |\x_0)$, in configuration space, for $E\gg (t-t_0)^{-1}$ gets strongly constrained to the classical path connecting the initial and final transverse positions, see Eq.~\eqref{eq:prop-G}, and leads to
\beq\label{eq:prop-G-eikonal}
(\x|\Gc_R(t,t_0)|\x_0)\simeq \Gc_0(\x-\x_0,t-t_0)\, V_R(t,t_0;[\x_{cl}(s)])  \,,
\eeq
where the classical trajectory is given by $\x_{\rm cl}(s) = \x_0 + \frac{s-t_0}{t-t_0}(\x-\x_0)$.
This corresponds to the product of a Wilson line, trailing the direction of the particle, times a vacuum propagator, see Eq.~\eqref{eq:G0}. Corrections to this limit can also be systematically be calculated \cite{Altinoluk_2014}. 
In the mixed representation, this leads to,
\begin{equation} \label{eikonal-prop-mom}
(\p|\Gc_R(t,t_0)|\p_0)\simeq(2 \pi)^{2} \delta(\boldsymbol{p}-\boldsymbol{p}_{0}) V_R\left(t, t_{0} ;\left[\boldsymbol{x}_{\mathrm{cl}}(s)=\bm n s\right]\right) \mathrm{e}^{-i \frac{\p^{2}}{2 E}\left(t-t_{0}\right)}\,.
\end{equation}
where $\n = \p/E$, see \cite{Mehtar_Tani_2018,Dom_nguez_2020}. The last term in this product is simply the Fourier transform of the vacuum propagator.

In detail, the $1 \to 2$ partonic processes we consider are: \texttt{1)} $\gamma \to q + \bar q$, \texttt{2)} $q \to q + g$, \texttt{3)} $g \to g + g$. These will, at most, involve correlators of 4, 6 and 8 Wilson lines (in the fundamental representation). We also write out the relevant correlators for $g \to q + \bar q$, but we do not explicitly evaluate the spectrum in this case.
All three processes consist of one (off-shell) particle\footnote{We will however only consider physical polarizations/spin states for the initial particle, since other contributions do not propagate.} traversing the medium splitting into two particles. While we derive formulas for a generic medium profile, our numerical calculations apply to a medium with constant density (aka the ``brick''), where the splitting can occur either inside the medium or outside.

As mentioned above, the first particle, with LC energy $E$, is produced at initial time $t_0=0$ and is propagating along the light-cone in the positive $z$ direction. It splits at times $t_1$ in the amplitude and $t_2$ in the complex conjugate amplitude, see Fig.~\ref{fig:pair-production-generic}.
The two daughter particles, which now carry LC energies $(1-z)E$ and $zE$, respectively, then propagate on the classical paths $\r_1(t)$ ($\r_{\bar 1}(t)$) and $\r_2(t)$ ($\r_{\bar 2}(t)$) in the amplitude (complex conjugate amplitude) to the end of the medium at $L$ . 
In the high-energy, eikonal approximation these paths are classical and are given by
\begin{equation}
\label{eq:classical-paths}
\begin{split}
\r_1(t) &= \bm n_1 (t-t_1) \,,\\
\r_2(t) &= \bm n_2 (t-t_1) \,,\\
\r_{\bar 1}(t) &= \bm n_1 (t-t_2) \,,\\
\r_{\bar 2}(t) &= \bm n_2 (t-t_2) \,,
\end{split}
\end{equation}
where $\bm n_1 \equiv \frac{\bm p_1}{zE}$ and $\bm n_2 \equiv \frac{\bm p_2}{(1-z)E}$ are the transverse velocity vectors. To slightly compress the notation we will usually refer to the coordinates as numbers, meaning that we will write $V(\r_1) \equiv V_1$ and $\gamma(\r_1-\r_{\bar 2}) \equiv \gamma_{1\bar 2}$, etc. 

Finally, in the harmonic approximation, we need the square of the differences of the transverse coordinates. Using the eikonal approximation this is
\begin{equation}
\label{eq:r-diff}
\begin{split}
    (\r_1-\r_2)^2&=(t-t_1)^2\theta^2 \,,\\
    (\r_{\bar 1}-\r_{\bar 2})^2&=(t-t_2)^2\theta^2 \,, \\
    (\r_1-\r_{\bar 1})^2&=(1-z)^2(t_2-t_1)^2\theta^2 \,, \\
    (\r_2-\r_{\bar 2})^2&=z^2(t_2-t_1)^2\theta^2 \,, \\
    (\r_1-\r_{\bar 2})^2&=(t-(1-z)t_1-zt_2)^2\theta^2 \,, \\
    (\r_{\bar 1}-\r_2)^2&=(t-z t_1-(1-z)t_2)^2\theta^2 \,,
\end{split}
\end{equation}
where we have assumed that the angle $\theta$ is small.

\section{Emission spectra}
\label{sec:split-processes}

In this section we will present the results for the in-medium emission spectra $\frac{\dd I}{\dd z \dd \theta}$ for the in-medium splitting processes. We refer to appendix \ref{appendix:calc-of-spectrums} for the details of the calculations. All of the Wilson line correlators in this section were calculated using the methods developed in Sec.~\ref{sec:Wilson}. For more details about the calculation of correlators of six and eight Wilson lines we refer to appendix \ref{sec:six-eight-lines}.

\begin{figure}[t!]
\centering
\includegraphics[width=0.6\textwidth]{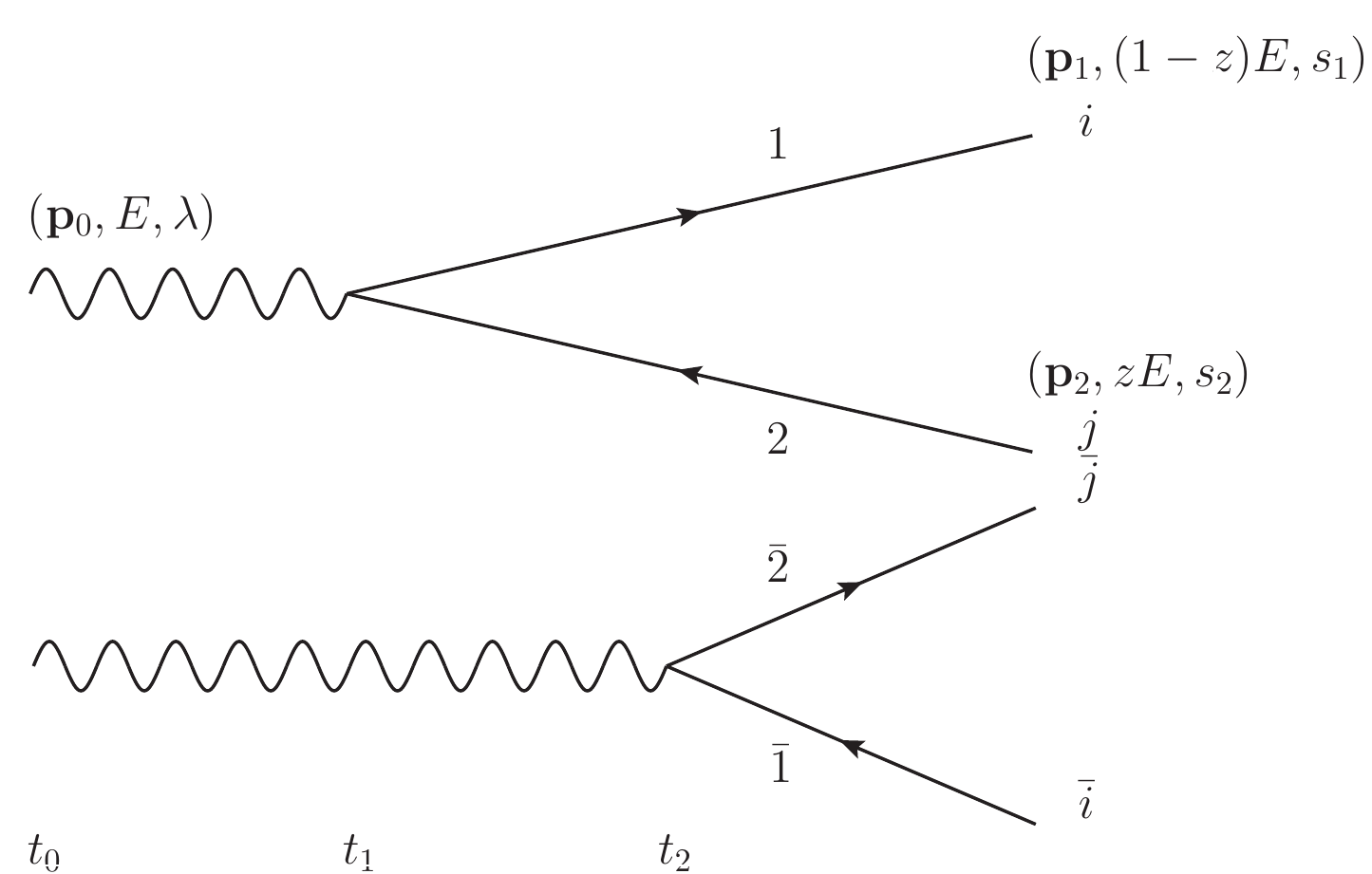} 
\caption{The process of a photon splitting to a quark-antiquark pair. The amplitude is on the top and the complex conjugate amplitude on the bottom. The splitting happens at time $t_1$ in the amplitude and at a later time $t_2$ in the complex conjugate amplitude.}
\label{fig:pair-production-generic}
\end{figure}

One can define the vacuum spectrum as
\begin{equation}\label{eq:vacuum-spectrum}
\frac{\mathrm{d} I^{\textrm{vac}}}{\mathrm{d} z \,\mathrm{d} \theta}=\frac{\alpha}{\pi} \frac{P(z)}{\theta}\,,
\end{equation}
where $\alpha$ can be $\alpha_{\textrm{em}}$ or $\alpha_s$ depending on the process, and $P(z)$ is the relevant Altarelli-Parisi splitting function. Then one can write the full spectrum on the form \cite{Dom_nguez_2020}
\begin{align}
\frac{\mathrm{d} I^{\textrm{full}}}{\mathrm{d} z \,\mathrm{d} \theta}&= \frac{\mathrm{d} I^{\textrm{vac}}}{\mathrm{d} z \,\mathrm{d} \theta}+\frac{\mathrm{d} I^{\textrm{med}}}{\mathrm{d} z \,\mathrm{d} \theta} \nn
&=\frac{\mathrm{d} I^{\textrm{vac}}}{\mathrm{d} z \,\mathrm{d} \theta}\left(1+F_\textrm{med}(z,\theta)\right)\,.
\end{align}
The term $F_\textrm{med}(z,\theta)$ contains the medium modification to the processes. For a generic medium profile the medium radiation reads
\begin{equation}
\label{eq:spectrum-generic-profile}
    \frac{\rmd I^{\rm med}}{\rmd z\, \rmd \theta} = \frac{\rmd I^{\rm vac}}{\rmd z\, \rmd \theta} \, 2 {\rm Re} \int_0^L \frac{\rmd t_1}{t_{\rm f}} \int_{t_1}^{L} \frac{\rmd t_2}{t_{\rm f}} \, \rme^{-i \frac{t_2-t_1}{t_{\rm f}}} \Cc^{(4)}(L,t_2)\Cc^{(3)}(t_2,t_1)\,,
\end{equation}
where, in the high-energy limit employed in this paper, the medium-induced spectrum is proportional to the vacuum spectrum. This proportionality does not \emph{a priori} hold in all the phase space, in particular whenever the transverse momentum in the splitting $k_\perp = z(1-z)E \theta$ is comparable to the transverse momentum accumulated in the medium $Q_s\sim \hat q L$ \cite{Blaizot_2013}.
Finally, the factors $\Cc^{(n)}(t_b,t_a)$ appearing in \eqref{eq:spectrum-generic-profile} are  $n$-particle correlators that have support during time $t_a < t <t_b$, and $t_{\rm f} =\frac{2}{z(1-z)E\theta^2}$ is the formation time of the process. The splitting process is illustrated in Fig.~\ref{fig:pair-production-generic}. 

For a medium with fixed density and extension $L$, we have $\hat q(t) = \hat q \, \Theta(L-t)$. In this case, the integrals over the emission times $t_1$ and $t_2$ in \eqref{eq:spectrum-generic-profile} can be split, so that 
\begin{align}
    \frac{\rmd I^{\rm in-in}}{\rmd z\, \rmd \theta} &= \frac{\rmd I^{\rm vac}}{\rmd z\, \rmd \theta} \, 2 {\rm Re} \int_0^L \frac{\rmd t_1}{t_{\rm f}} \int_{t_1}^{L} \frac{\rmd t_2}{t_{\rm f}} \, \rme^{-i \frac{t_2-t_1}{t_{\rm f}}} \Cc^{(4)}(L,t_2) \Cc^{(3)}(t_2,t_1)\,, \\
    \frac{\rmd I^{\rm in-out}}{\rmd z\, \rmd \theta} &= \frac{\rmd I^{\rm vac}}{\rmd z\, \rmd \theta} \, 2 {\rm Im} \int_0^L \frac{\rmd t_1}{t_{\rm f}} \, \rme^{-i \frac{L-t_1}{t_{\rm f}}} \Cc^{(3)}(L,t_1)\,,
\end{align}
where $\rmd N^{\rm med}/(\rmd z\rmd \theta) = \rmd N^{\rm in-in}/(\rmd z\rmd \theta)+\rmd N^{\rm in-out}/(\rmd z\rmd \theta)$.
Taking into account that the Wilson line correlators are real the medium modification term can be written \cite{Dom_nguez_2020}
\begin{equation}\label{eq:fmed-final}
    F_{\textrm{med}}=2\int_0^L \frac{\dd t_1}{t_{\rm f}}
    \left[\int_{t_1}^L \frac{\dd t_2}{t_{\rm f}}\cos\left(\frac{t_2-t_1}{t_{\rm f}}\right)\Cc^{(4)}(L,t_2) \Cc^{(3)}(t_2,t_1)
    -\sin\left(\frac{L-t_1}{t_{\rm f}}\right)\Cc^{(3)}(L,t_1)\right]\,.
\end{equation}
We now will discuss three concrete cases that are relevant for jet quenching phenomenology. We will compute the double-differential spectrum for a wide range of LC energy sharing fraction $z$ and angles $\theta$ to map the regions where medium-induced corrections appear, as quantified by the factor $F_{\rm med}(z,\theta)$. Our focus here is to provide a test bed for evaluating precisely the multi-Wilson line correlators appearing in \eqref{eq:fmed-final}, and we will therefore not worry about the validity of the eikonal approximation \eqref{eq:prop-G-eikonal} of the splitting products. Including non-eikonal corrections on the particle trajectories will be postponed to future work.


\subsection{Derivation of the splitting functions}
\label{sec:split-processes-derivation}

\paragraph{Photon splitting}
We will start with the case of a photon splitting into a quark-antiquark pair, i.e. $\gamma \to q + \bar q$. Due to the least number of fundamental Wilson lines, this is the simplest process to analyze. We will therefore treat it in more detail, taking the advantage to discuss the relevant medium and jet scales appearing in the calculation.

In this case, the vacuum emission spectrum is given by \eqref{eq:vacuum-spectrum} with the QED coupling constant $\alpha_{\rm em}$ and the Altarelli-Parisi splitting function being $P_{q \gamma}(z)= n_f N_c[z^2+(1-z)^2]$, where $n_f$ is the number of active flavors. Furthermore, the correlator $\Cc^{(3)}$ reduces to an effective two-point function because the photon does not carry color charge. We have
\begin{align}
    \label{eq:c4-photon-splitting}
    \cC_{q \gamma}^{(4)}(L,t_2) &= \frac{1}{N_c} \langle \tr [V_{1} V_{2}^{\dagger} V_{\bar{2}} V_{\bar{1}}^{\dagger}]\rangle \,,\\
    \label{eq:c3-photon-splitting}
    \cC_{q \gamma}^{(3)}(t_2,t_1) &= \frac{1}{N_c} \tr \langle V_1 V_2^\dagger \rangle \,,
\end{align}
where the time extension of each of the medium-averaged color correlators on the right hand side is implied by the time argument on left hand side of the equation.

The correlator of two Wilson lines, which in this case corresponds to $\cC_{q \gamma}^{(3)}(t,t_1) = \Sc_{12}(t,t_1)$, is generally referred to as a dipole correlator, and is known to be
\begin{equation}
\Sc_{12}(t,t_1) \equiv \frac{1}{N_c}\langle \tr[V_1 V_2^\d] \rangle= \rme^{-C_F \int_{t_1}^{t} \dd s \, n(s) \sigma(\r)} \,,
\end{equation}
where $\r = \r_1 - \r_2$ is the difference of transverse positions of the two Wilson lines. For a fixed separation, i.e. $\r = $ const., in the HO approximation and in a medium with constant density, it simply reads $\Sc_{12}(t,t_1) = \rme^{- \frac{1}{4} \hat q (t-t_1) \, \r^2}$, where, as a reminder, we have denoted $\hat q \equiv \hat q_F$. However, for the kinematics we consider, see Eq.~\eqref{eq:r-diff}, this becomes
\begin{equation}
\label{eq:c3-photon-splitting-2}
    \Sc_{12}(t,t_1) = \rme^{-\frac{1}{12} \hat q (t-t_1)^3 \theta^2 } \,.
\end{equation}
Then, assuming that $t=t_2$ and $t_2 - t_1 \sim t_{\rm f}$, we find
\begin{equation}
    \Sc_{12} \approx \rme^{- \frac{2}{3}\frac{\hat q}{\omega^3 \theta^4} } \,,
\end{equation}
with $\omega = z(1-z)E$. This implies that medium modifications appear, in this term, whenever $\omega^3 \theta^4 \lesssim \hat q$.

The correlator of four Wilson lines $\cC^{(4)}(t,t_2)$, referred to as the quadrupole (in the fundamental representation), can only be calculated numerically at finite-$N_c$. We will later show how this can be achieved through the differential equation Eq.~\eqref{diff-matrix-new}. In the large-$N_c$ limit, however, it can be calculated analytically through the simplified differential equation Eq.~\eqref{diff-matrix-largeNc}. There are only two ways of connecting the Wilson lines at the final time (their connection at initial time is given by the vacuum splitting process). We can therefore define $C_{1 \bar 2}(t,t_2) \equiv \langle\tr[ V_{1} V_2^{\dagger}]\tr[V_{\bar2} V_{\bar{1}}^{\dagger}] \rangle$ and 
$ C_{\bar 2 1}(t,t_2) \equiv \langle \tr [V_{1} V_{2}^{\dagger} V_{\bar{2}} V_{\bar{1}}^{\dagger}]\rangle$, note the absence of explicit normalization factors at this stage. In the large-$N_c$ approximation, the first of these correlators reads simply,
\begin{equation}\label{Q2-largeNc}
\frac{1}{N_c^2}  C_{1\bar 2}(t,t_2) \simeq \rme^{-\frac{1}{12}\hat q\theta ^2 \left[(t-t_2)^3+(t-t_1)^3-\tau^3\right]}\,,
\end{equation}
where $\tau \equiv t_2-t_1$, and the dependence on $t_1$ appears as a consequence of the fixed trajectories.
Here we have used the eikonal \eqref{eq:classical-paths} and harmonic oscillator approximations \eqref{eq:harmonic-pot}. Similarly, at large-$N_c$, the second correlator is
\begin{align}
\label{eq:Q1-consistent-Nc}
\frac{1}{N_c}  C_{\bar 2 1}(t,t_2) &\simeq \rme^{-\frac{1}{4}\hat q\theta ^2 \xi(t-t_2)\tau^2}\nn
&-\frac12 \hat q \theta^2 z(1-z) \tau^2 \int^t_{t_2} \mathrm{d}s \,
e^{-\frac{1}{4}\hat q \theta^2\xi(t-s)\tau^2}
e^{-\frac{1}{12} \hat q\theta ^2\left[(s-t_2)^{3}+(s-t_1)^{3}-\tau^{3}\right]}\,,
\end{align}
where we defined $\xi \equiv z^2+(1-z)^2$. Only the latter of these correlators appears in the spectrum, cf. Eq.~\eqref{eq:c4-photon-splitting}, but we include both for completeness.\footnote{In \cite{Dom_nguez_2020} $C_{\bar 2 1}$ is also calculated in the large-$N_c$ limit. In their Eq. (29) they get the same as \eqref{eq:Q1-consistent-Nc}, except they lack the factor of $1/2$ in front of the second term.} Assuming the dominance of the first term in \eqref{eq:Q1-consistent-Nc}, setting $t=L$ and assuming that $L \gg t_{\rm f}$, we find that 
\begin{equation}
    \frac{1}{N_c}  C_{\bar 2 1} \approx \rme^{- \frac{1}{6} \frac{\hat qL}{(\omega \theta)^2}} \,,
\end{equation}
where we put $\xi \approx 2/3$.
The factor in the exponential becomes large whenever $\omega \theta < \sqrt{\hat qL}$. This factor is related to momentum broadening of the quark and anti-quark after they have been produced.

Let us compare the two conditions when exponential suppression arise either in the dipole $\Sc_{12}(t_2,t_1)$ or quadrupole $ C_{\bar 2 1}(L,t_2)$. For a fixed energy $\omega$, the two conditions are equal at the critical angle
\begin{equation}
\theta_c \sim \left(\hat q L^3 \right)^{-1/2} \,.
\end{equation}
Let us also define the characteristic energies $\omega_{\rm d} = (\hat q/\theta^4)^{1/3}$ and $\omega_{\rm broad} = \sqrt{\hat q L}/\theta$.
At large angles $\theta > \theta_c$, 
the condition from the dipole starts affecting soft gluon emissions, i.e. $\omega_{\rm d} < \omega_{\rm broad}$.
This reflects the length-dependence color coherence. On the one hand, the dipole, which has support only during the formation time $t_{\rm f} \lesssim L$, needs a large angle to resolve the two particles within that time scale. On the other hand, the quadrupole, which extends up to $L$, will ultimately resolve even narrower configurations.

We also plot the dependence on the latest time of both $\cC_{q \gamma}^{(3)}(t,t_1)$ and $\cC_{q \gamma}^{(4)}(t,t_2)$ in Fig.~\ref{fig:C3-C4-q-gamma}, keeping $t_1 = 0.3$ fm fixed, in the case of the dipole, and both $t_2 =1$ fm and $t_1 = 0.3$ fm fixed, in the case of the quadrupole. The other parameters are chosen as $\hat q = 1.5$ GeV$^2/$fm, $\theta =0.5$ and $z = 0.5$. We notice the fast decay of the dipole, that goes like $\sim \rme^{-t^3}$ according to \eqref{eq:c3-photon-splitting-2}, compared to the exponential decay of the quadrupole, i.e. $\sim \rme^{-t}$, at large times. Finally, we notice that the large-$N_c$ approximation to the full quadrupole, given in Eq.~\eqref{eq:Q1-consistent-Nc}, is very good up very late times.

\paragraph{Quark-gluon splitting}
Next we consider the slightly more complicated problem of a quark-gluon splitting. 
This was also outlined in \cite{Dom_nguez_2020}, but not calculated explicitly.
For this process, the vacuum emission spectrum is given by \eqref{eq:vacuum-spectrum}, with the QCD coupling constant $\alpha_s$ and the Altarelli-Parisi splitting function $P_{gq}(z)= C_F \frac{1+(1-z)^2}{z}$. The four- and three-point functions read
\begin{align}
    \label{eq:c4-quark-gluon}
    \cC_{gq}^{(4)}(L,t_2) &= \frac{1}{N_c^2-1} \left\langle \tr[V_{\bar 1}^\dagger V_1V_2^\dagger V_{\bar 2}] \tr[V_{\bar 2}^\dagger V_2]-\frac{1}{N_c}\tr[V_{\bar 1}^\dagger V_1]\right\rangle  \,,\\
    \label{eq:c3-quark-gluon}
    \cC_{gq}^{(3)}(t_2,t_1) &= \frac{1}{N_c^2-1} \left\langle \tr[V_2^\dagger V_1]\tr[V_0^\dagger V_2]-\frac{1}{N_c}\tr[V_0^\dagger V_1] \right\rangle\,.
\end{align}
The emission spectrum is composed of correlators of two, four and six Wilson lines. The three-point function can be solved exactly, see \eqref{3-Wilson-line-easy}, resulting in
\begin{align}
\label{eq:c3-quark-splitting}
    \cC_{gq}^{(3)}(t_2,t_1) &= \rme^{-\frac12 \int_{t_1}^{t_2} \dd s\, n(s)[N_c(\sigma_{02}+\sigma_{12})-\frac{1}{N_c}\sigma_{01}]} \nn
    &= \rme^{-\frac{1}{12} \qhat(t_2-t_1)^3 \theta^2 \left(1+z^2+\frac{2z}{N_c^2-1}\right)}\,.
\end{align}
This expression is very similar to the dipole term in Eq.~\eqref{eq:c3-photon-splitting-2} and the same scale analysis applies.

The four-point correlator involving six and two Wilson lines can only be calculated numerically at finite $N_c$. In the large-$N_c$ limit, the former can be calculated analytically, and reads
\begin{align} \label{3-lines-largeNc}
&\frac{1}{N_c^2}\langle\tr[V_1 V_2^\d V_{\bar 2} V_{\bar 1}^\d]\tr [V_2 V_{\bar 2}^\d]\rangle \simeq 
\rme^{-\frac14\qhat\theta^2(t-t_2)(t_2-t_1)^2(1-2z+3z^2)} \nn
&\times\left(1-\frac12\qhat\theta^2z(1-z)(t_2-t_1)^2 
\int_{t_2}^t \dd s \, 
\rme^{-\frac{1}{12}\qhat\theta^2 \left[(s-t_2)^2(2s-3t_1+t_2)+6z(1-z)(s-t_2)(t_2-t_1)^2\right]}\right)\,.
\end{align}
Once again, the first term in the correlator above has a form very similar to the four-point function relevant for photon splitting, see Eq.~\eqref{eq:Q1-consistent-Nc}.   

\paragraph{Gluon-gluon splitting}
The last process of interest is the case of a gluon splitting into two other gluons.
This process was discussed quite extensively in \cite{Blaizot_2013}. For this process, the vacuum emission spectrum is given by \eqref{eq:vacuum-spectrum} with the QCD coupling constant $\alpha_s$ and the Altarelli-Parisi splitting function $P_{gg}(z)= 2N_c\big[z(1-z)+\frac{1-z}{z}+\frac{z}{1-z}\big]$. In this case the 4- and 3-point functions read
\begin{align}
    \label{eq:c4-gluon-gluon-fund}
    \cC_{gg}^{(4)}(L,t_2) &= \frac{1}{N_c(N_c^2-1)} \left\langle \tr[V_1V_{\bar 1}^\dagger] \tr[V_2 V_{\bar 2}^\dagger V_{\bar 1}V_1^\dagger] \tr[V_{\bar 2}V_2^\dagger]  -\tr[V_1 V_{\bar 1}^\dagger V_2V_{\bar2}^\dagger V_{\bar1} V_1^\dagger V_{\bar 2}V_2^\dagger] \right\rangle \,,\\
    \label{eq:c3-gluon-gluon-fund}
    \cC_{gg}^{(3)}(t_2,t_1) &= \frac{1}{N_c(N_c^2-1)} \left\langle \tr[V_1 V_2^\dagger] \tr[V_0 V_1^\dagger] \tr[V_2 V_0^\dagger] - \tr[V_1 V_2^\dagger V_0 V_1^\dagger V_2 V_0^\dagger]\right\rangle\,.
\end{align}
When cast as correlators of Wilson lines in the fundamental representation, the $\cC^{(4)}_{gg}$ involves 8-point correlators, which is the largest number we will calculate in detail.

The 3-point function can be solved exactly, either by the differential equation \eqref{master-diff-eq} or by writing it in terms of adjoint Wilson lines \eqref{ggg-Wilson-lines-t2t1}. In the end, the result reads 
\begin{align}
\label{eq:c3-gluon-gluon-adjoint}
\cC_{gg}^{(3)}(t_2,t_1) &= \rme^{-\frac{N_c}{2}\int_{t_1}^{t_2}\dd t\, n(t)[\sigma_{01}+\sigma_{02}+\sigma_{12}]}\nn
&= \rme^{-\frac{1}{12}\qhat(t_2-t_1)^3\theta^2\frac{N_c}{C_F}(1-z+z^2)}\,.
\end{align}
Note the similarity to the previous results, see Eqs.~\eqref{eq:c3-photon-splitting-2} and \eqref{eq:c3-quark-splitting}.

The 4-point function consists of two different correlators of eight Wilson lines. They can be calculated through the differential equation in Eq.~\eqref{master-diff-eq}. Interestingly, the four-point function $\cC^{(4)}_{gg}$ involves a eight-point correlator, see the second term in \eqref{eq:c4-gluon-gluon-fund}, which cannot be reduced further in the large-$N_c$ approximation. This can nevertheless still be exactly solved in the large-$N_c$ approximation, which we present in the figures below, but the expression is too long to extract any meaningful approximation. Anticipating the numerical results, we can mention that it is for this correlator that the large-$N_c$ approximation gives the biggest deviations with respect to the exact result.

\paragraph{Gluon-quark splitting}
We now consider a gluon that splits into a quark-antiquark pair. The Altarelli-Parisi splitting function is $P_{q g}(z)= n_f T_R[z^2+(1-z)^2]$ 
and the correlators read
\begin{align}
    \label{eq:c4-gluon-quark}
    \cC_{q g}^{(4)}(L,t_2) &= \frac{1}{N_c} \left\langle \tr[V_1 V_{2}^\dagger V_{\bar 2} V_{\bar 1}^\dagger] -\frac{1}{N_c}\tr[V_1 V_{\bar 1}^\dagger] \tr[V_{\bar 2} V_2^\dagger] \right\rangle \,,\\
    \label{eq:c3-gluon-quark}
    \cC_{q g}^{(3)}(t_2,t_1) &= \frac{1}{N_c^2-1} \left\langle \tr[V_1 V_0^\dagger] \tr[V_0 V_2^\dagger] - \frac{1}{N_c}\tr[V_1 V_2^\dagger]\right\rangle\,.    
\end{align}
Since these expressions involve only quadrupoles and dipoles, that were previously encountered and analyzed in detail above, we will not present further results for this splitting process.


\subsection{Numerical results}
Here we present the numerical calculations of the results from the previous section. We focus first on the details of the three- and four-point functions for each of the three splitting processes, and proceed with calculating the double-differential spectrum in the momentum sharing fraction $z$ and angle $\theta$.

\begin{figure}[t!]
\centering
    \begin{subfigure}[b]{0.45\textwidth}
        \centering
        \includegraphics[width=\textwidth]{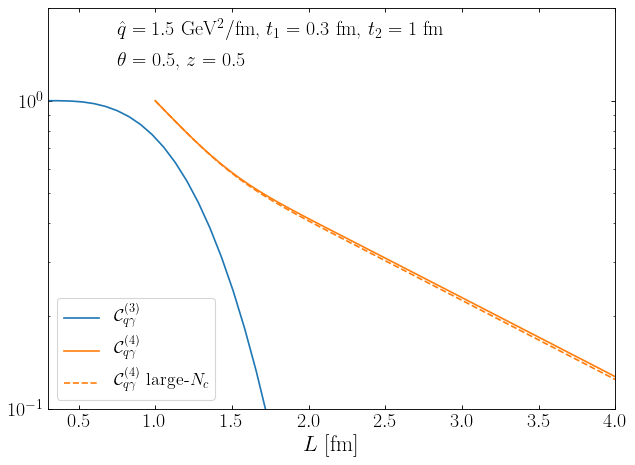}
        \caption{Photon splitting.}
        \label{fig:C3-C4-q-gamma}
    \end{subfigure}
    \hfill
    \begin{subfigure}[b]{0.45\textwidth}
        \centering
        \includegraphics[width=\textwidth]{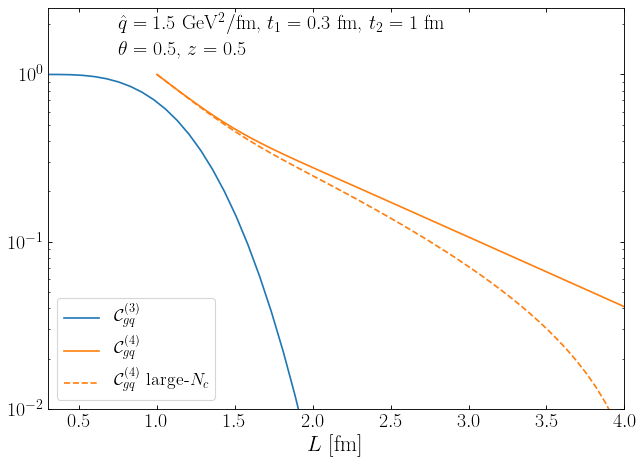}
        \caption{Quark-gluon splitting.}
        \label{fig:C3-C4-g-q}
    \end{subfigure}
    \hfill
    \begin{subfigure}[b]{0.45\textwidth}
        \centering
        \includegraphics[width=\textwidth]{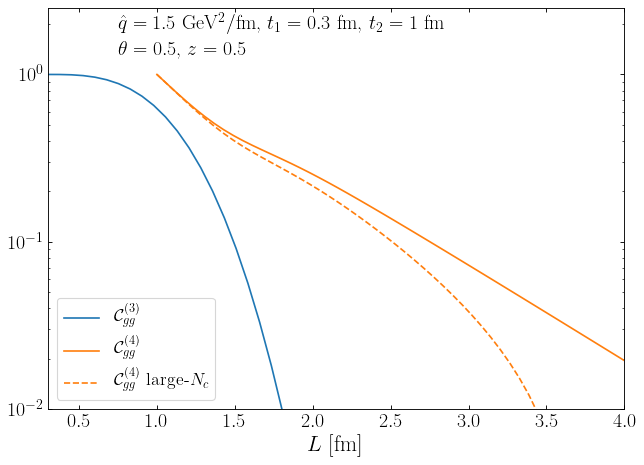}
        \caption{Gluon-gluon splitting.}
        \label{fig:C3-C4-g-g}
    \end{subfigure}
\caption{The time evolution of $\cC^{(3)}(L,t_1)$ and $\cC^{(4)}(L,t_2)$ for the three processes. For $\cC^{(4)}(L,t_2)$ both the exact and large-$N_c$ versions are plotted.}
\label{fig:numerics-correlators}
\end{figure}
In Fig.~\ref{fig:numerics-correlators}, we show how $\cC^{(3)}_{ij}(t,t_1)$, with blue, solid curves, and $\cC^{(4)}_{ij}(t,t_2)$, with orange, solid curves, for the three processes evolve with time. For the four-point functions, we also plot the large-$N_c$ approximation with orange, dashed curves. We fix both $t_1 = 0.3$ fm and $t_2 =1$ fm and plot for the latest time $t=L$. For the other parameters we choose $\hat q = 1.5$ GeV$^2/$fm, $\theta =0.5$ and $z = 0.5$.

While this approximation turns out work extremely well for the photon splitting, see Fig.~\ref{fig:C3-C4-q-gamma}, we note that it has a more limited range of applicability for both the quark-gluon, see Fig.~\ref{fig:C3-C4-g-q}, and gluon-gluon, see Fig.~\ref{fig:C3-C4-g-g}, splitting processes, respectively. 
In all of the cases the exact value is slightly higher than the approximate one. As we derived analytically, the $\cC^{(3)}_{ij}$ terms all decay as $\sim \rme^{-\qhat(t-t_1)^3\tau^2 \theta^2 f(z) }$, where $f(z)$ is a process dependent regular function. The $\cC^{(4)}_{ij}$ terms are more complicated, especially at early times where all terms contribute, but at late times the dominant contribution comes from $\sim \rme^{-\qhat(t-t_2)\theta^2}$.

\begin{figure}
    \centering
    \begin{subfigure}[b]{0.45\textwidth}
       \centering
       \includegraphics[width=\textwidth]{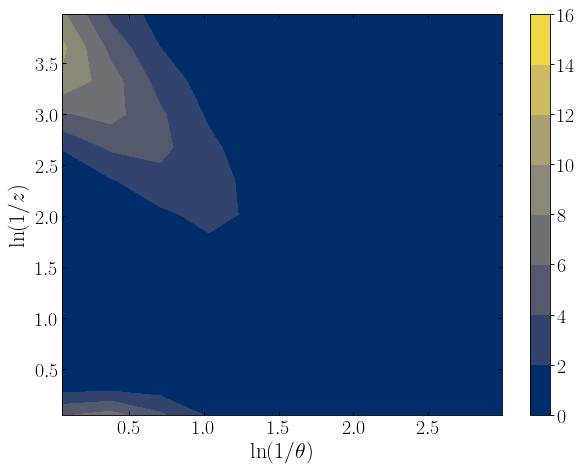}
       \caption{Photon splitting.}
       \label{fig:spectrum-photon}
    \end{subfigure}
    \hfill
    \begin{subfigure}[b]{0.45\textwidth}
       \centering
       \includegraphics[width=\textwidth]{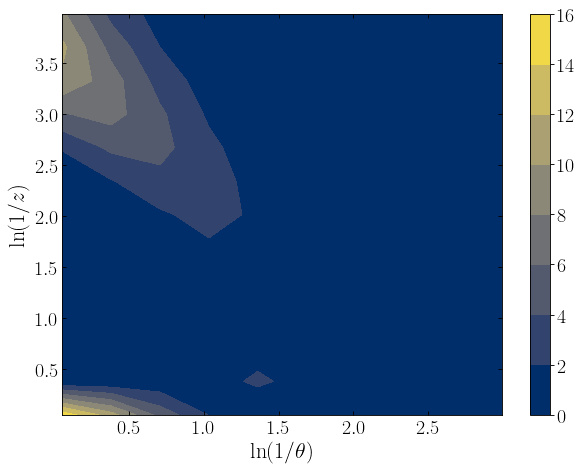}
       \caption{Quark-gluon splitting.}
       \label{fig:spectrum-qg}
    \end{subfigure}
    \hfill
    \begin{subfigure}[b]{0.45\textwidth}
       \centering
       \includegraphics[width=\textwidth]{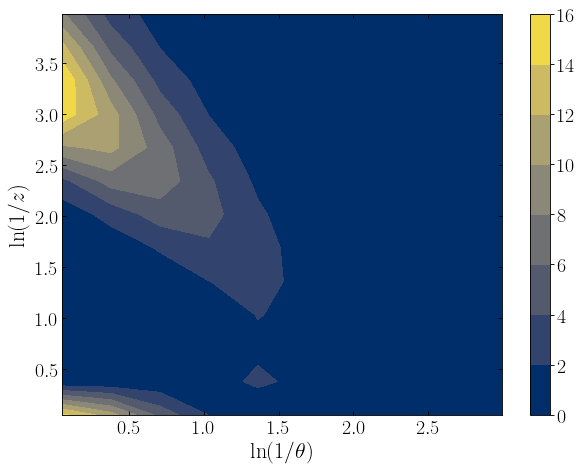}
       \caption{Gluon-gluon splitting.}
       \label{fig:spectrum-gg}
    \end{subfigure}
    \caption{The medium modification factor $F_\textrm{med}(z,\theta)$ for three splitting processes as a function of $\theta$ and $z$ with $L=2$ fm and $E=100$ GeV at finite $N_c$.}
    \label{fig:spectrum-pairprod}
\end{figure}
The ratio of double-differential in-medium to vacuum spectrum reveals the medium modification factor $F_{\rm med}(z,\theta) = \rmd I^{\rm med}/(\rmd z \rmd \theta) \big/ \rmd I^{\rm vac}/(\rmd z \rmd  \theta)$. We plot this factor, calculated at finite $N_c$, for the three processes in Fig.~\ref{fig:spectrum-pairprod}. These results have been obtained for the medium parameters $\hat q = 1.5$ GeV$^2$/fm and $L = 2$ fm and an energy of the initial particle, before splitting, of $E = 100$ GeV. 

As one can see from Fig.~\ref{fig:spectrum-pairprod}, the medium modification factor $F_\textrm{med}(z,\theta)$ has roughly the same characteristic shape for all three processes. The medium modifications appear at large angles $\theta > \theta_c$, in between the characteristic lines $\omega^3 \theta^3 < \hat q$ and $\omega^2 \theta^2 < \hat q L$ which we have identified for the three- and four-point functions in Sec.~\ref{sec:split-processes-derivation}. This corresponds to formation times smaller than the medium length, $t_{\rm f} < L$. In fact, we can recast these conditions in terms of the formation time of the process, namely $t_{\rm f} < t_{\rm d}$ and $t_{\rm f} < t_{\rm broad}$, where
\begin{equation}
    t_{\rm d} \sim \left(\frac{1}{\hat q \theta^2} \right)^{1/3} \,, \qquad \text{and } \qquad t_{\rm broad} \sim \left(\frac{1}{\hat q \theta^2 L} \right)^{1/2} \,.
\end{equation}
The modifications appear for the range of formation times $t_{\rm broad} < t_{\rm f} < t_{\rm d}$ and $\theta > \theta_c$ \cite{Dom_nguez_2020}.
There also seems to be a trend that both the magnitude and the region of the modifications grow with the number of Wilson lines. This can be traced back to the finite terms, $f(z)$, in the exponents that modify the scaling behavior. Naively, we would expect the relevant jet quenching parameter to be roughly a factor $N_c/C_F \approx 2$ larger for gluon splitting than for the photon.

\begin{figure}
    \centering
    \begin{subfigure}[b]{0.45\textwidth}
       \centering
       \includegraphics[width=\textwidth]{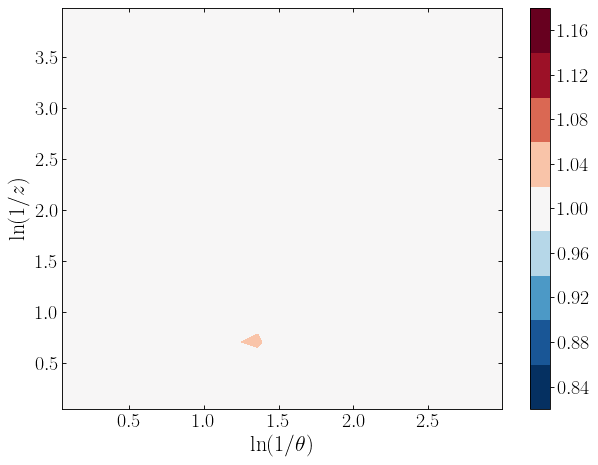}
       \caption{Photon splitting.}
       \label{fig:lund-photon}
    \end{subfigure}
    \hfill
    \begin{subfigure}[b]{0.45\textwidth}
       \centering
       \includegraphics[width=\textwidth]{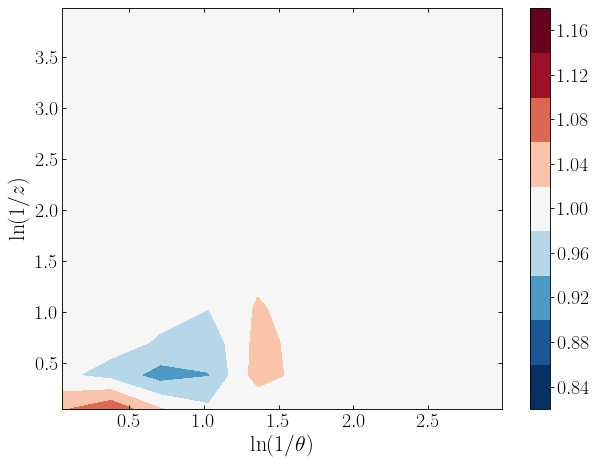}
       \caption{Quark-gluon splitting.}
       \label{fig:lund-qg}
    \end{subfigure}
    \hfill
    \begin{subfigure}[b]{0.45\textwidth}
       \centering
       \includegraphics[width=\textwidth]{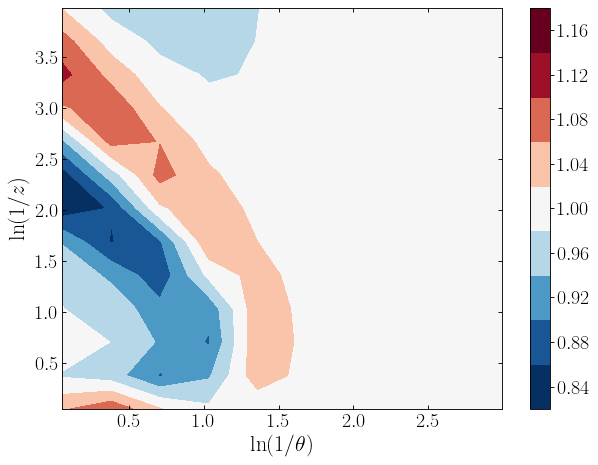}
       \caption{Gluon-gluon splitting.}
       \label{fig:lund-gg}
    \end{subfigure}
    \caption{The ratio $F_{\rm med}(z,\theta)|_{{\rm large-}N_c}/F_{\rm med}(z,\theta)$ for three splitting processes as a function of $\theta$ and $z$ with $L=2$ fm and $E=100$ GeV.}
    \label{fig:lund-ratio}
\end{figure}
Our main focus in this work is to highlight the differences between the finite-$N_c$ results versus their large-$N_c$ approximated counterparts. 
To illustrate this we have plotted the ratio of the exact and large-$N_c$ medium modification factors, i.e. $F_{\rm med}(z,\theta)|_{{\rm large-}N_c}/F_{\rm med}(z,\theta)$ 
in Fig.~\ref{fig:lund-ratio}. 
The difference between the exact and approximate result is small in the whole phase space in the photon splitting case, where there is a correlator of four Wilson lines. However, in the cases of quark-gluon and especially gluon-gluon splitting, which contain correlators of six and eight Wilson lines, the error can be relatively big, maximally of the order of 16\% in case of the latter process. This is the reflection of the behavior observed previously in Fig.~\ref{fig:numerics-correlators}. From these calculations it seems like the more complicated color structure, the bigger the error is by using the large-$N_c$ approximation. Once again, the error becomes most sizable at relatively large in-medium formation times, i.e. $t_{\rm f} \sim t_{\rm broad}$ and $t_{\rm f} \sim t_{\rm d}$, but at the same time $t_{\rm f} < L$. This is most clearly seen in the gluon-gluon splitting, cf. Fig.~\ref{fig:lund-gg}. Finally, we note that the finite-$N_c$ corrections come as a modulation along the previously established scaling lines which hints that such corrections could perhaps be absorbed into an \emph{effective} jet quenching parameter.

To summarize, we have calculated the the double-differential spectrum $\frac{\dd I}{\dd z \dd \theta}$ for three different splitting processes, and shown that the resulting expressions factorize into three- and four-point functions that contain medium-averaged products of 2, 4, 6 and 8 fundamental Wilson lines. In the coming Sec.~\ref{sec:Wilson} we will detail how these are calculated. Strikingly, the three- and four-point functions all take a very similar scaling form as was derived analytically exactly, for the former, and in the large-$N_c$ approximation, for the latter. This corresponds to the identification of two characteristic time-scales in the medium, related to broadening along the length of the medium, $t_{\rm broad}$, and decoherence during the formation of the splitting, $t_{\rm d}$. These were identified first in \cite{Dom_nguez_2020} for the photon splitting process, and we have here extended their validity to all other splitting QCD processes. Finally, we have seen that finite-$N_c$ corrections play an increasingly important role the bigger the total color charge involved in the splitting process.

\section{Calculating Wilson line correlators}\label{sec:Wilson}

In this section we will present our method for calculating Wilson line correlators. As an illustration we will first show how it is done in the simple case of four Wilson lines in the fundamental representation. Thereafter this process will be generalized to an arbitrary number of Wilson lines.
\subsection{Four Wilson lines}\label{sec:four-lines}
\begin{figure}[t!]
\centering
\includegraphics[width=0.9\textwidth]{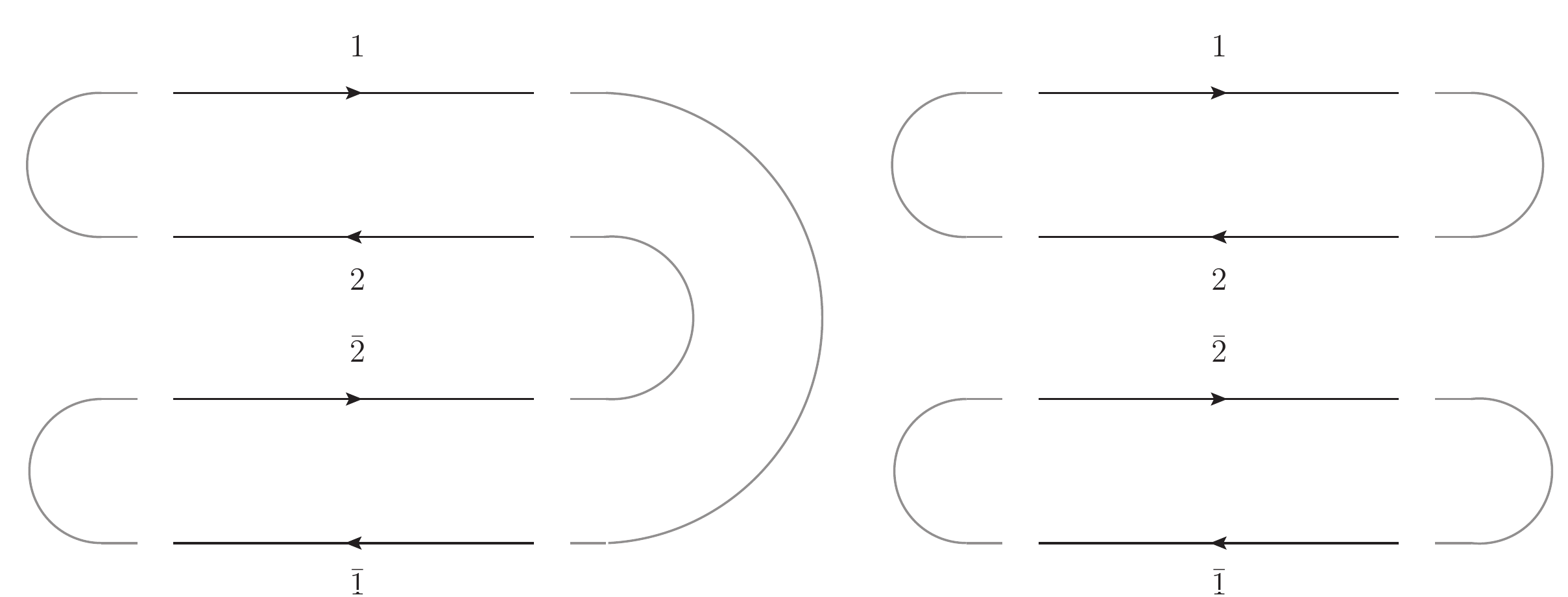} 
\caption{The two possible ways of color connecting the four Wilson lines. On the left is $C_{\bar 2 1} \equiv \langle \tr [V_{1} V_{2}^{\dagger} V_{\bar{2}} V_{\bar{1}}^{\dagger}]\rangle$, while on the right is $ C_{1 \bar 2} \equiv \langle\tr[ V_{1} V_2^{\dagger}]\tr[V_{\bar2} V_{\bar{1}}^{\dagger}] \rangle$, both  going from times $t_2$ to an arbitrary time $t$. The grey lines at the beginning and end indicate the colour connections.}
\label{fig:4-lines}
\end{figure}

The simplest Wilson line correlator comes from the pair production process \eqref{eq:c4-photon-splitting}, where there is a trace of four Wilson lines $\langle \tr [V_{1} V_{2}^{\dagger} V_{\bar{2}} V_{\bar{1}}^{\dagger}]\rangle$. In this section, we will show how to derive a system of differential equations to calculate this. Let the Wilson lines have support from $t_0$ to some arbitrary time $t+\epsilon$. Then, following \cite{Kovner_2001}, we expand them between $t$ and $t+\epsilon$ to get
\begin{align} 
\label{Wilson-expansion}
&V(t+\epsilon,t_0;\mathbf{r})
= V(t+\epsilon,t;\mathbf{r})V(t,t_0;\mathbf{r}) \nn
&=\left(1+i g \int^{t+\epsilon}_t \rmd s\, A^{a}(s, \mathbf{r}) t^{a}-\frac{g^2}{2!}  \int^{t+\epsilon}_t \rmd s \int^{t+\epsilon}_t \rmd s'\, A^{a}(s, \mathbf{r})A^{b}(s', \mathbf{r}') t^{a} t^{b} + \mathcal{O}(\epsilon^2)\right)\nn
&\times V(t,t_0;\mathbf{r}),
\end{align}
where we have kept some of the color indices implicit.
All four Wilson lines are expanded in this manner. We end up with having to take the medium average of the integrals over two medium fields, traced over the relevant color indices, which is dealt in the following way
\begin{align} 
\label{integrals-medium-avg}
\int^{t+\epsilon}_t \rmd s \int^{t+\epsilon}_t \rmd s' \left\langle \, A^{a}(s, \mathbf{r})A^{b}(s', \mathbf{r}') \tr[t^{a} t^{b}]\right\rangle 
&=\int^{t+\epsilon}_t \dd s \,n(s) \gamma(\r -\r') t_{ij}^{a} t_{ji}^{a}\nn
&\simeq \epsilon \,C_F n(t)\gamma(\r -\r') \,,
\end{align}
where in the first step we applied the medium average \eqref{eq:med-avg}.
Then, keeping terms up to the first order of $\epsilon$ this becomes
\begin{align}
\langle \tr [V_{1} V_{2}^{\dagger} V_{\bar{2}} V_{\bar{1}}^{\dagger}]\rangle _{(t+\epsilon)} 
&= \left(1+\epsilon g^2 n(t) C_F[\gamma_{1\bar 1} + \gamma_{2\bar 2}-2 \gamma_0]\right) 
\langle \tr [V_{1} V_{2}^{\dagger} V_{\bar{2}} V_{\bar{1}}^{\dagger}]\rangle _{(t)}\nn
&- \epsilon g^2 n(t)[\gamma_{1\bar{2}} - \gamma_{12} - \gamma_{\bar 1\bar 2} + \gamma_{2\bar{1}}] 
\langle \tr[t^a V_{1} V_{2}^{\dagger}  t^a V_{\bar{2}} V_{\bar{1}}^{\dagger}]\rangle _{(t)} \,.
\end{align}
Using the Fierz identity
\begin{equation} \label{fierz}
t_{i j}^{a} t_{k l}^{a}=\frac{1}{2}\left(\delta_{i l} \delta_{j k}-\frac{1}{N_c} \delta_{i j} \delta_{k l}\right)\, ,
\end{equation}
this results in the differential equation
\begin{align}\label{Q1-diff-eq}
&\frac{\dd}{\dd t} \langle \tr [V_{1} V_{2}^{\dagger} V_{\bar{2}} V_{\bar{1}}^{\dagger}]\rangle _{(t)} 
= \lim_{\epsilon\to 0}\frac{\langle \tr [V_{1} V_{2}^{\dagger} V_{\bar{2}} V_{\bar{1}}^{\dagger}]\rangle _{(t+\epsilon)}- \langle \tr [V_{1} V_{2}^{\dagger} V_{\bar{2}} V_{\bar{1}}^{\dagger}]\rangle _{(t)}}{\epsilon} \nn
&= g^2 n(t) \left[C_F(\gamma_{1\bar 1} + \gamma_{2\bar 2}-2 \gamma_0)
+\frac{1}{2 N_c}(\gamma_{1\bar{2}} - \gamma_{12} - \gamma_{\bar 1\bar 2} + \gamma_{2\bar{1}})\right] \langle \tr [V_{1} V_{2}^{\dagger} V_{\bar{2}} V_{\bar{1}}^{\dagger}]\rangle _{(t)} \nn
&-\frac12 g^2 n(t) (\gamma_{1\bar{2}} - \gamma_{12} - \gamma_{\bar 1\bar 2} + \gamma_{2\bar{1}}) \langle\tr[ V_{1} V_2^{\dagger}]\tr[V_{\bar2} V_{\bar{1}}^{\dagger}] \rangle_{(t)} \,.
\end{align}
It is evident that the original term mixes with another four-point correlator, given in the term on the last line. To understand this, let us look closer at the term $\langle \tr [V_{1} V_{2}^{\dagger} V_{\bar{2}} V_{\bar{1}}^{\dagger}]\rangle$. Since the process is happening in the medium the quarks and antiquarks can at any time exchange gluons, so their color is continuously rotating. In the case of four Wilson lines there are two possible ways of connecting the color at time $t$ to ensure color conservation, namely as shown in Fig.~\ref{fig:4-lines}. The second way is exactly the term $\langle\tr[ V_{1} V_2^{\dagger}]\tr[V_{\bar2} V_{\bar{1}}^{\dagger}] \rangle$ that appeared in equation \eqref{Q1-diff-eq}. 
The inclusion of this term in the differential equation \eqref{Q1-diff-eq} just represents the possibility for color rotation to happen at each time.

To continue one can find a complementary differential equation for $\langle\tr[ V_{1} V_2^{\dagger}]\tr[V_{\bar2} V_{\bar{1}}^{\dagger}] \rangle$ and see if we can find a solution for the set. Going through the same procedure as above gives
\begin{align}\label{Q2-diff-eq}
&\frac{\dd}{\dd t} \langle\tr[ V_{1} V_2^{\dagger}]\tr[V_{\bar2} V_{\bar{1}}^{\dagger}] \rangle_{(t)} \nn
&= g^2 n(t) \left[C_F(\gamma_{1 2} + \gamma_{\bar1\bar{2}}-2 \gamma_0 )
+\frac{1}{2 N_c}(\gamma_{1\bar{2}} - \gamma_{1\bar1} - \gamma_{2\bar{2}} + \gamma_{\bar{1}2})\right] \langle\tr[ V_{1} V_2^{\dagger}]\tr[V_{\bar2} V_{\bar{1}}^{\dagger}] \rangle_{(t)} \nn
&-\frac12 g^2 n(t) (\gamma_{1\bar{2}} - \gamma_{1\bar1} - \gamma_{2\bar{2}} + \gamma_{\bar{1}2})\langle \tr [V_{1} V_{2}^{\dagger} V_{\bar{2}} V_{\bar{1}}^{\dagger}]\rangle_{(t)} \,.
\end{align}
We now have a set of two coupled differential equations. To save space the following notation will be used $C_{1 \bar 2}(t) \equiv \langle\tr[ V_{1} V_2^{\dagger}]\tr[V_{\bar2} V_{\bar{1}}^{\dagger}] \rangle_{(t)}$ and $C_{\bar 2 1}(t) \equiv \langle \tr [V_{1} V_{2}^{\dagger} V_{\bar{2}} V_{\bar{1}}^{\dagger}]\rangle _{(t)}$. This notation warrants some more explanation. Both of these expressions are composed of the two pairs of Wilson lines, namely $V_{1} V_2^{\dagger}$ and $V_{\bar2} V_{\bar{1}}^{\dagger}$. The only difference is how to connect them. The two subscripts in the $C$'s tell which Wilson line that comes immediately after the two pairs. So $C_{\bar 2 1}$ means that $V_{1} V_2^{\dagger}$ is connected to $V_{\bar 2}$ and $V_{\bar2} V_{\bar{1}}^{\dagger}$ connects to $V_1$. The result is $\langle \tr [V_{1} V_{2}^{\dagger} V_{\bar{2}} V_{\bar{1}}^{\dagger}]\rangle$. This notation might seem overly complicated, but it will prove to be useful when considering more than four Wilson lines.

The two differential equations \eqref{Q1-diff-eq} and \eqref{Q2-diff-eq} can be gathered into the following system,
\begin{equation}
\label{eq:diff-matrix-exact}
\frac{\dd}{\dd t}
\begin{bmatrix}
C_{1\bar 2}(t)\\
C_{\bar 2 1}(t)
\end{bmatrix}
= - \frac{n(t)}{2} \mathbb{M}
\begin{bmatrix}
C_{1\bar 2}(t)\\
C_{\bar 2 1}(t)
\end{bmatrix}
\,,
\end{equation}
where the evolution matrix takes the following form,
\begin{equation}\label{eq:evolution-matrix}
    \mathbb{M} = 
\begin{bmatrix}
2 C_F(\sigma_{12} + \sigma_{\bar{2}\,\bar{1}})+\frac{1}{N_c}\Sigma_1
& -\Sigma_1 \\
-\Sigma_2
& 2C_F(\sigma_{1\bar1} + \sigma_{\bar{2}2})+\frac{1}{N_c}\Sigma_2
\end{bmatrix} \,.
\end{equation}
Here we have used Eq.~\eqref{eq:sigma-gamma} to define $\sigma_{12} = \sigma(\r_1 - \r_2)$, and introduced
\begin{align}
\Sigma_1 &\equiv \sigma_{1\bar{2}} + \sigma_{2\bar{1}} - \sigma_{1\bar{1}} - \sigma_{2\bar{2}}  \nn
 \Sigma_2 &\equiv \sigma_{1\bar{2}} + \sigma_{\bar{1}2}- \sigma_{12} - \sigma_{\bar1\bar{2}}\,.
\end{align}
To proceed, we employ the the harmonic approximation \eqref{eq:harmonic-pot}. For the eikonal, straight-line trajectories, given in Eqs.~\eqref{eq:classical-paths}, the evolution matrix becomes
\begin{equation}
\label{diff-matrix-new}
- \frac{n(t)}{2}\mathbb{M} = - \frac{\hat{q}\theta^2}{4 C_F}
\begin{bmatrix}
 C_F[(t-t_1)^2+(t-t_2)^2]-\frac{1}{N_c}(t-t_1)(t-t_2)& 
- (t-t_1)(t-t_2) \\
  z(1-z)\tau^2& 
C_F\tau^2\xi-\frac{1}{N_c}z(1-z)\tau^2
\end{bmatrix}
\,,
\end{equation}
where we have defined $\tau \equiv t_2-t_1$, $\xi = z^2+(1-z)^2$ and assumed that the angle between the two particles $\theta$ is small. Unfortunately, since the matrix elements depend on time in our setup, we can only solve this system of differential equations exactly by using numerical methods. 

The authors of \cite{Dom_nguez_2020} calculated the four-point function $\langle \tr [V_{1} V_{2}^{\dagger} V_{\bar{2}} V_{\bar{1}}^{\dagger}]\rangle$ in the large-$N_c$ limit, which is interesting to compare with our results. This example is illustrative of the general structure of the hierarchy between the different correlators, and we will therefore go through it in detail. To take the large-$N_c$ limit you start the system of differential equations \eqref{eq:diff-matrix-exact} and count the powers of $N_c$ in each term in the evolution matrix and the vector of correlators, taking into account that $C_{1\bar 2} \sim N_c^2$ and $C_{\bar 2 1} \sim N_c^1$. In this limit we also have $C_F \sim N_c/2$. The terms on the right-hand side of \eqref{eq:diff-matrix-exact} then have the following powers of $N_c$,
\begin{align}
\label{eq:largeNc-scaling-quadrupole}
\begin{bmatrix}
\mathcal{O}(N_c^0) + \mathcal{O}(N_c^{-2})& 
\mathcal{O}(N_c^{-1}) \\
\mathcal{O}(N_c^{-1}) &
\mathcal{O}(N_c^0) + \mathcal{O}(N_c^{-2})
\end{bmatrix}
\begin{bmatrix}
\mathcal{O}(N_c^2)\\
\mathcal{O}(N_c^1)
\end{bmatrix} 
&\xrightarrow[]{\text{large}-N_c} 
\begin{bmatrix}
    \mathcal{O}(N_c^0) & 
    0 \\
    \mathcal{O}(N_c^{-1}) &
    \mathcal{O}(N_c^0) 
\end{bmatrix}
\begin{bmatrix}
\mathcal{O}(N_c^2)\\
\mathcal{O}(N_c^1)
\end{bmatrix}
\,.
\end{align}
The large-$N_c$ approximation amounts to dropping all the terms in the matrix that are not scaling with the same power of $N_c$ as the original vector, given by the second term in \eqref{eq:largeNc-scaling-quadrupole}. We see that the the next-to-leading power of $N_c$ turns out to be a factor $N_c^{-2}$ smaller compared to the leading terms. This scaling has also been corroborated generally for $n$-line correlators in Sec.~\ref{sec:Wilson-general}.

Hence, employing the large-$N_c$ approximation leads to the simplified system of equations
\begin{equation} \label{diff-matrix-largeNc}
\frac{\dd}{\dd t}
\begin{bmatrix}
C_{1\bar 2}(t)\\
C_{\bar 2 1}(t)
\end{bmatrix}
\simeq - \frac{\hat{q}\theta^2}{4 N_c}
\begin{bmatrix}
N_c [(t-t_1)^2+(t-t_2)^2]& 
0\\
2  z(1-z)\tau^2 & 
N_c \tau^2\xi
\end{bmatrix}
\begin{bmatrix}
C_{1\bar 2}(t)\\
C_{\bar 2 1}(t)
\end{bmatrix}
\,.
\end{equation}
Now it is evident that the differential equation for $C_{1\bar 2}$ is separable and can be solved easily, which means that $C_{\bar 2 1}$ also can be solved. This leads to the equations \eqref{Q2-largeNc} and \eqref{eq:Q1-consistent-Nc}. 
The physical picture of this differential equation is quite transparent. The correlator of the two particles (described by two lines in the amplitude and two lines in the complex conjugate amplitude) can be in either of the states shown in Fig.~\ref{fig:4-lines}, and there is a possibility of exchanging a gluon and transferring from one state to the other. This is encoded in the off-diagonal terms in the matrix \eqref{eq:evolution-matrix}, and is associated with a factor of $\sim \sigma$, which scales as $N_c^{-1}$. Say you start in the state $C_{1\bar 2}$ shown on the right in figure \ref{fig:4-lines}, scaling as $N_c^2$. If you exchange a gluon you pick up a factor $N_c^{-1}$ from the $\sigma$, and go to the state $C_{\bar 2 1}$, which is a single trace correlator that scales as $N_c^1$, so in total this transition is associated with a factor $N_c^0$. This is a factor $N_c^{-2}$ smaller compared to the starting point so it can safely be dropped in the large-$N_c$ limit. However, starting with $C_{\bar 2 1}$ and going to $C_{1 \bar 2 }$ you go from a state that scales as $N_c^1$ to one scaling as $N_c^2$, but you lose a power of $N_c$ from the $\sigma$, so in this case the end result has the same $N_c$ scaling as the starting point. Hence, in the large-$N_c$ limit you can drop the upper right term in the matrix, but must keep the lower left one, see Sec.~\ref{sec:Wilson-general} for a general argument for $n$-point correlators.

\begin{figure}[t!]
\begin{center}
\includegraphics[width=0.8\textwidth]{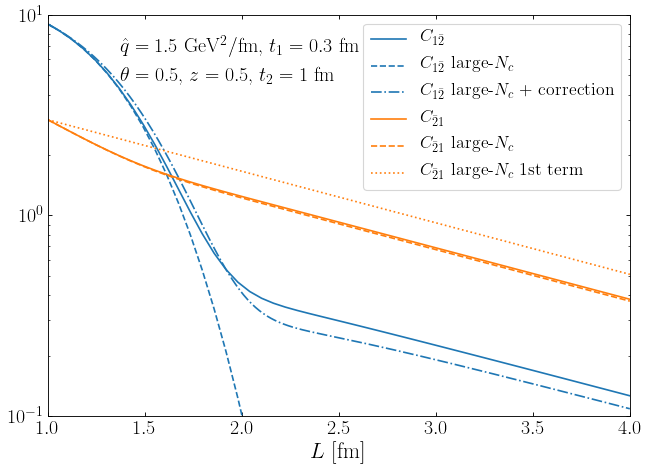} 
\end{center}
\caption{The exact and large-$N_c$ version of $C_{1 \bar 2}(t) = \langle\tr[ V_{1} V_2^{\dagger}]\tr[V_{\bar2} V_{\bar{1}}^{\dagger}] \rangle$ (blue, solid and blue, dashed lines, respectively) and $C_{\bar 2 1}(t) = \langle \tr [V_{1} V_{2}^{\dagger} V_{\bar{2}} V_{\bar{1}}^{\dagger}]\rangle$ (orange, solid and orange, dashed lines, respectively). We also plot only the leading, diagonal term of the large-$N_c$ approximation of $C_{\bar 2 1}$ (orange, dotted line) which exhibits the correct large-time asymptotic behavior.}
\label{fig:4-Wilson-lines}
\end{figure}
The solutions to \eqref{diff-matrix-new} and \eqref{diff-matrix-largeNc} were plotted in Fig.~\ref{fig:4-Wilson-lines} (solid and dashed lines, respectively). For this particular case, the agreement between the large-$N_c$ approximation and the exact, finite-$N_c$ result is strikingly good for the $C_{\bar 2 1}$ correlator. At late times, we observe an exponential suppression, $\propto \rme^{-t}$, with a slope that is in good agreement with the first term of Eq.~\eqref{eq:Q1-consistent-Nc}. At early times, there is an interplay between $C_{\bar 2 1 }$ and $C_{1 \bar 2}$ that leads to a more rapid decrease initially. This is however well captured by the large-$N_c$ approximation, given by both terms in Eq.~\eqref{eq:Q1-consistent-Nc}. 

The $C_{1\bar 2}$ correlator is described well within the large-$N_c$ approximation at early times. However, at late times it exhibits a long tail that is not captured within this approximation. This can be remedied by including sub-leading corrections in color.

Sub-leading corrections in color can be incorporated to improve on the sometimes crude large-$N_c$ calculation above. To do this write the full correlators as the sum of their large-$N_c$ versions calculated through \eqref{diff-matrix-largeNc} and some smaller correction term,
\begin{equation}
\label{eq:largeNc-correction-1}
    \C = \C^{(0)} + \C^{(1)}\,,
\end{equation}
where $\C = \big(C_{1 \bar 2}, C_{\bar 2 1}\big)^\intercal$ is a vector of the correlators in question, so that $\C^{(1)}$ is a factor $\mathcal{O}(N_c^{-2})$ smaller than $\C^{(0)}$. We can also write the matrix $\mathbb{M}$ in a form that isolates the large-$N_c$ terms from the finite-$N_c$ corrections, i.e.
\begin{equation}
\label{eq:largeNc-correction-2}
    \mathbb{M} = \mathbb{M}^{(0)} + \mathbb{M}^{\rm corr.} \,,
\end{equation}
where the first term strictly corresponds to the leading terms in the large-$N_c$ limit. In our example above, we find that 
\begin{equation}
- \frac{n(t)}{2}\mathbb{M}^{(0)} = - \frac{\hat{q}\theta^2}{4}
\begin{bmatrix}
(t-t_1)^2+(t-t_2)^2 & 
0 \\
\frac{2}{N_c} z(1-z)\tau^2& 
\tau^2\xi
\end{bmatrix}
\,,
\end{equation}
while 
\begin{equation}
- \frac{n(t)}{2}\mathbb{M}^{\rm corr.} \simeq  \frac{\hat{q}\theta^2}{2 N_c^2}
\begin{bmatrix}
(t-t_1)(t-t_2)& 
N_c (t-t_1)(t-t_2) \\
-\frac{1}{N_c}z(1-z)\tau^2 & 
z(1-z)\tau^2
\end{bmatrix}
\,,
\end{equation}
where we expanded the correction matrix to find the leading terms in $N_c$. It can be confirmed that the overall correction to both correlators is of the order $N_c^{-2}$.

The correlators at leading color, i.e. $\C^{(0)}$, are known.  They solve the simplified set of equations $\rmd \C^{(0)}(t)/\rmd t = -\frac{n(t)}{2} \mathbb{M}^{(0)} \C^{(0)}(t)$, and are given explicitly in \eqref{Q2-largeNc} and \eqref{eq:Q1-consistent-Nc}. This can now be used to calculate the color sub-leading contributions $\C^{(1)}$. Simply plugging this into the full differential equation \eqref{eq:diff-matrix-exact} results in the following differential equation for the first correction
\begin{align} \label{diff-matrix-correction}
\frac{\dd}{\dd t}
\C^{(1)}(t)
\simeq & -\frac{n(t)}{2}\mathbb{M}^{(0)} \C^{(1)}(t) - \frac{n(t)}{2} \mathbb{M}^{\rm corr} \C^{(0)}(t)
\,,
\end{align}
where we have neglected terms that are even more sub-leading, i.e. resulting from $\mathbb{M}^{\rm corr.}\C^{(1)}$.
This is an nonhomogeneous version of the large-$N_c$ system of differential equations \eqref{diff-matrix-largeNc}, and can also be solved exactly. As an example the first correction to $C^1_{1 \bar 2 }(t)$ is
\begin{align}
    C^{(1)}_{1 \bar 2 }(t) &=\frac{\qhat \theta^2}{2N_c^2}\int_{t_2}^t \rmd s \,(s-t_1)(s-t_2) \left[ C^{(0)}_{1\bar 2}(s) + N_c C^{(0)}_{\bar 2 1}(s) \right]\nn
    &\times \rme^{\frac{\qhat \theta^2}{12}[(t-t_1)^3-(s-t_1)^3+(t-t_2)^3-(s-t_2)^3]}\,.
\end{align}
The first correction contains $C^{(0)}_{\bar 2 1}(t)$, given in \eqref{eq:Q1-consistent-Nc} which as can be seen in Fig.~\ref{fig:4-Wilson-lines} has a linear tail at long times. One would therefore expect that this correction will rectify the difference between the exact calculation and the large-$N_c$ version of $C_{1 \bar 2 }(t)$ at long times which can be seen in the same plot. On Fig.~\ref{fig:4-Wilson-lines}, we have plotted this correction, and it is indeed clear that it contains this linear tail. It is also worth noticing that the $N_c$-scaling of the correction is $C^{(1)}_{1 \bar 2 }\sim N_c^0$, since there is an $N_c^{-1}$ in the pre-factor and $C^{(0)}_{\bar 2 1} \sim N_c^1$. As expected the correction is lower by a factor $N_c^{-2}$ compared to the large-$N_c$ result.

It is possible to calculate higher order corrections going as $N_c^{-4}$, $N_c^{-6}$ etc. compared to the large-$N_c$ expression using the same technique recursively. 

\subsection{General method for Wilson line correlators}
\label{sec:Wilson-general}

In Sec.~\ref{sec:split-processes}, we showed that doing similar calculations starting with a quark or a gluon emitting a gluon leads to correlators of six and eight fundamental Wilson lines, respectively. We will now generalize the procedure demonstrated in the preceding section and develop a method of calculating correlators of an arbitrary number of fundamental Wilson lines. To be more precise we get systems of differential equations like in \eqref{eq:diff-matrix-exact}, and will show how to easily calculate all the matrix elements in the $K!\times K!$ matrices. The system can then be solved numerically or, as we will see, analytically in the large-$N_c$ limit.

The correlators of six and eight Wilson lines that appeared in Sec.~\ref{sec:split-processes} are
\begin{itemize}
    \item $\langle \tr[V_{\bar 1}^\dagger V_1V_2^\dagger V_{\bar 2}] \tr[V_{\bar 2}^\dagger V_2]\rangle$,
    \item $\langle \tr[V_1V_{\bar 1}^\dagger]\tr[V_2 V_{\bar 2}^\dagger V_{\bar 1}V_1^\dagger]\tr[V_{\bar 2}V_2^\dagger]\rangle$,
    \item and $\langle\tr[V_1 V_{\bar 1}^\dagger V_2V_{\bar2}^\dagger V_{\bar1} V_1^\dagger V_{\bar 2}V_2^\dagger]\rangle$.
\end{itemize}  
Note that one can divide the correlators of these Wilson lines into pairs on the form $[V_n V_{\bar m}^\dagger]_{i_n j_m}$ times some Kronecker deltas that connect the indices. 

To start, consider the special case of calculating a correlator involving $K$ pairs of a Wilson line in the amplitude times the same Wilson line in the complex conjugate amplitude
\begin{equation}\label{K-lines}
\langle[V_{1} V_{\bar 1}^\dagger]_{i_1j_1}[V_{2} V_{\bar 2}^\dagger]_{i_2j_2} \dots [V_{K} V_{\bar K}^\dagger]_{i_K j_K}\rangle=\langle\prod_{n=1}^K  [V_{n} V_{\bar n}^\dagger]_{i_nj_n}\rangle \,.
\end{equation}
This is very useful to consider, even though none of the correlators mentioned above are of this exact form. The reason is that in this form all of the formulas derived in this section become much nicer. In addition, it is easy to generalize this to include all cases simply by changing the labels of the Wilson lines in \eqref{K-lines} to whatever is needed in the specific problem at hand. For example, choosing $K=3$ and changing labels $(1,\bar 1,2,\bar 2,3,\bar 3)\to(1,2,\bar 2,\bar 1,2,\bar 2)$ gives the structure needed in \eqref{eq:c4-quark-gluon}, while $K=4$ and changing labels $(1,\bar1,2,\bar2,3,\bar 3,4, \bar 4) \to (1,\bar1,2,\bar2,\bar 1,1,\bar 2,2)$ reproduces the correlators in \eqref{eq:c4-gluon-gluon-fund}. So even though it seems we are calculating a special case, simply changing the labels in the equations in this section leads to all possible cases. 

To compress the notation a bit we will write the k'th instance of a Wilson line pair as 
\begin{equation}
W^k_{i_k j_k} \equiv [V_{k} V_{\bar k}^\dagger]_{i_kj_k}\,.
\end{equation}
It is possible to generalize the method of reaching a system  of differential equations showed in the previous section to an arbitrary number $K$ pairs of Wilson lines. The steps are outlined in App.~\ref{appendix:proof-matrix-eq}. This procedure leads to the differential equation,
\begin{align} \label{K-Wilson-lines}
&\frac{2 N_c}{g^2}\frac{\rmd}{\rmd t}\left\langle\prod_{n=1}^K  W^n_{i_n j_n} \right\rangle  \nn
&= n(t)\left[ \sum_{k=1}^{K-1} \sum_{l>k}^K(\gamma_{k l}+\gamma_{\bar k  \bar l}-\gamma_{k  \bar l}-\gamma_{\bar k  l})-\sum_{k=1}^K\gamma_{k  \bar k}-K(N_c^2-1)\gamma_0\right] \left\langle\prod_{n=1}^K W^n_{i_n j_n} \right\rangle \nn
&+n(t)\sum_{k=1}^K \left[ \gamma_{k \bar k} \left\langle \tr( W^k) \delta_{i_k j_k} \left(\prod_{n \neq k}^K W^n_{i_n j_n}\right) \right \rangle \right]\nn
&+n(t)\sum_{k=1}^{K-1} \sum_{l>k}^K \Bigg\langle \left(\gamma_{k  \bar l}\delta_{i_k j_l}[ W^l  W^k]_{i_l j_k}+\gamma_{\bar k  l}\delta_{i_lj_k}[ W^k  W^l]_{i_k j_l}-\gamma_{k l}W^k_{i_l j_k}W^l_{i_k j_l}-\gamma_{\bar k  \bar l} W^k_{i_k j_l} W^l_{i_l j_k} \right) \nn
&\quad \times \prod_{n \neq k, n \neq l}^K W^n_{i_n j_n} \Bigg \rangle\,.
\end{align}
One can see that the term on the first line has the same index structure as the original, while the subsequent lines contain mixing terms. Notice that, in the mixing terms, only at most two $W$'s change place. The rest stay the same as before. 

The above equation is a step in the right direction. It makes it possible to quite easily project out all the different differential equations by contraction with the product of $K$ Kronecker deltas. For example starting with \eqref{K-Wilson-lines} and projecting out with $\delta_{j_1i_1}\delta_{j_2i_2}\dots \delta_{j_Ki_K}$ turns it into a differential equation for $\frac{\rmd}{\rmd t}\langle tr[W^1]\tr[W^2]\dots\tr[W^K]\rangle$, while $\delta_{j_1i_2}\delta_{j_2i_3}\dots \delta_{j_Ki_1}$ leads to $\frac{\dd}{\dd t}\langle \tr[W^1W^2\dots W^K]\rangle$. We will denote these two possibilities by  $C_{12\dots K}\equiv \langle\tr[W^1]\tr[W^2]\dots\tr[W^K]\rangle$ and $C_{23\dots K1}\equiv \langle\tr[W^1W^2\dots W^K]\rangle$. The general version of this is $C_{m_1m_2\dots m_K}$, where $m_1m_2\dots m_K$ is one of the $K!$ permutations of the numbers between $1$ and $K$. The idea behind this notation is that $W^1$ is connected to $W^{m_1}$, $W^2$ is connected to $W^{m_2}$ etc.\footnote{One final example to clarify the notation can, for instance, be the correlator $C_{213\ldots K}\equiv \langle \tr [W^1 W^2] \tr[W^3] \ldots \tr [W^K]\rangle$.} 

Although it is possible to use \eqref{K-Wilson-lines} to project out all the necessary differential equations, there are actually $K!$ such projections, which quickly becomes a huge number. It would be much preferable to write this system in matrix form, like in Eq.~\eqref{eq:diff-matrix-exact}. Making use of the notation we described above we want to write the system of differential equations for $K$ pairs of Wilson lines as
\begin{equation} \label{any-N-differential-system}
\frac{\dd}{\dd t}C_{m_1m_2\dots m_K} = -\frac12 n(t) \sum_{p_1p_2\dots p_K} \mathbb{M}_{m_1m_2\dots m_K}^{p_1p_2\dots p_K} C_{p_1p_2\dots p_K}\,,
\end{equation}
where $p_1p_2\dots p_K$ also is one of the $K!$ permutations of $12\dots K$.

Starting from \eqref{K-Wilson-lines}, one can deduce the general form of the matrix elements $\mathbb{M}_{m_1m_2\dots m_K}^{p_1p_2\dots p_K}$. For details on how this is done, we refer to App.~\ref{appendix:proof-matrix-eq}. Fortunately, most of the matrix elements are zero, and those that are not have quite simple expressions. The $K!$ diagonal entries are
\begin{align} \label{a-diagonal}
\mathbb{M}_{m_1m_2\dots m_K}^{m_1m_2\dots m_K}&=N_c\sum_{k=1}^K \sigma_{\bar k m_k}+\frac{1}{N_c} \underbrace{\sum_{k=1}^K \sum_{l>k}^K(\sigma_{k l}+\sigma_{\bar k \bar l}-\sigma_{k \bar l}-\sigma_{\bar k  l})-\frac{1}{N_c}\sum_{k=1}^K\sigma_{k \bar k}}_{A_K}
\end{align}
Note here that only the first sum depends on the exact permutation we use. The two latter sums are independent of this, and are common to all the diagonal terms, so we call it $A_K$. The only other non-zero matrix elements $\mathbb{M}_{m_1m_2\dots m_K}^{p_1p_2\dots p_K}$ are those where $p_1p_2\dots p_K$ is just $m_1m_2\dots m_K$, but with exactly two entries swapped places. If our original sequence is $m_1m_2\dots m_i \dots m_j \dots m_K$, and its entries in positions $i$ and $j$ have changed places it becomes $m_1m_2\dots m_j \dots m_i \dots m_K$. Then we get $K!\frac{K(K-1)}{2}$ entries of the form
\begin{equation}\label{a-non-diagonal}
\mathbb{M}_{m_1m_2\dots m_i \dots m_j \dots m_K}^{m_1m_2\dots m_j \dots m_i \dots m_K}=\sigma_{\bar i m_j}+\sigma_{m_i\bar j }-\sigma_{m_i m_j}-\sigma_{\bar i \bar j}\,.
\end{equation}
Finally, we have
\begin{equation}
\mathbb{M}_{m_1m_2\dots m_K}^{p_1p_2\dots p_K}=0\,,
\end{equation}
for $p_1p_2\dots p_K$ being any other permutation of $m_1 m_2 \dots m_K$. This means that out of the $K!^2$ matrix elements, only $\frac12 K!(K^2-K+2)$ are non-zero. These are given by the relatively simple formulas \eqref{a-diagonal} and \eqref{a-non-diagonal}. Putting it all together this becomes
\begin{align} \label{master-diff-eq}
\frac{\dd}{\dd t}C_{m_1m_2\dots m_i \dots m_j \dots m_K} = 
&-\frac12 n(t) \big(N_c\sum_{k=1}^K \sigma_{\bar k m_k}+\frac{1}{N_c}A_K\big)C_{m_1m_2\dots m_i \dots m_j \dots m_K} \nn
&-\frac12 n(t)\sum_{i=1}^{K-1}\sum_{j>i}^K (\sigma_{\bar i m_j}+\sigma_{m_i\bar j }-\sigma_{m_i m_j}-\sigma_{\bar i \bar j})C_{m_1m_2\dots m_j \dots m_i \dots m_K} \,.
\end{align}
Of course, for all differential equations you need to specify some initial conditions. It is clear from the definition of the Wilson line \eqref{Wilson-definition} that $V_{ij}(t_0,t_0)=\delta_{ij}$. The trace of this is $\tr V(t_0,t_0)$=$N_c$. This means that the initial condition of a Wilson line correlator is $N_c$ to the power of traces it contains. A few illustrative examples of this are $C_{12\dots K}= \langle\tr[W^1]\tr[W^2]\dots\tr[W^K]\rangle \sim N_c^K$, $C_{213\dots K}= \langle\tr[W^1W^2]\tr[W^3]\dots\tr[W^K]\rangle \sim N_c^{K-1}$ and $C_{23\dots K1}= \langle\tr[W^1W^2\dots W^K]\rangle \sim N_c^1$.

To better understand what this system of differential equations looks like, it is useful to view it in matrix form. Generally, there will be several correlators that go as the same power of $N_c$. It is useful to gather these in vectors $\bm{C^M}$, where the superscript $M$ is meant to indicate that this scales as $N_c^M$. Then, Eq.~\eqref{master-diff-eq} can be represented as
\begin{equation}\label{matrix-representation}
\frac{\dd}{\dd t}
\begin{bmatrix}
C^K\\
\bm{C^{K-1}}\\
\bm{C^{K-2}}\\
\vdots\\
\bm{C^{2}}\\
\bm{C^1}
\end{bmatrix}
\sim
\left(
\textrm{diag}\left(N_c \sigma + \frac{1}{N_c}\sigma,\dots,N_c \sigma + \frac{1}{N_c}\sigma\right)
+
\begin{bmatrix}
0 & \bm{\sigma} & \bm0 & \dots  & \dots   & \bm 0\\
\bm{\sigma} & \bm{0} & \bm{\sigma} & \bm{0} & \dots & \bm 0\\
\bm0 & \bm{\sigma} & \bm{0} & \bm{\sigma} & \bm{0} & \bm 0\\
\vdots & \vdots & \vdots & \vdots & \vdots & \vdots \\
\bm{0} & \dots & \bm{0} & \bm{\sigma} & \bm{0} & \bm{\sigma}\\
\bm{0} & \dots & \dots & \bm{0} & \bm{\sigma}&\bm0\\
\end{bmatrix}
\right)
\begin{bmatrix}
C^K\\
\bm{C^{K-1}}\\
\bm{C^{K-2}}\\
\vdots\\
\bm{C^{2}}\\
\bm{C^1}
\end{bmatrix}
\,.
\end{equation}
The first matrix contains the diagonal elements, written in detail in \eqref{a-diagonal}. The second matrix represents the non-diagonal elements, and $\bm{\sigma}$ is a block containing non-zero elements, which we get from Eq.~\eqref{a-non-diagonal}.

\subsection{Wilson line correlators in the large-$N_c$ limit}\label{sec:Wilson-line-largeNc}

The system of differential equations given by Eq.~\eqref{master-diff-eq} is to our knowledge not possible to solve analytically in the case where $\sigma$ is a function of time, so we have to turn to numerical techniques. However, in the large-$N_c$ limit, the system simplifies in a way that makes it possible to solve it exactly. This can be seen from the matrix representation in Eq.~\eqref{matrix-representation}. Since $\sigma \sim N_c^{-1}$, the diagonal matrix elements go as $\sim N_c^0+ N_c^{-2}$, while the non-diagonal ones go as $\sim N_c^{-1}$. Multiplying in the vector on the end and representing every term by its $N_c$ scaling this becomes
\begin{equation}
\frac{\dd}{\dd t}
\begin{bmatrix}
C^K\\
\bm{C^{K-1}}\\
\bm{C^{K-2}}\\
\vdots\\
\bm{C^{2}}\\
\bm{C^1}
\end{bmatrix}
\sim
\begin{bmatrix}
(N_c^0 + N_c^{-2})C^K\\
(\bm{N_c^{0}}+ \bm{N_c^{-2}})\bm{C^{K-1}}\\
(\bm{N_c^{0}}+ \bm{N_c^{-2}})\bm{C^{K-2}}\\
\vdots\\
(\bm{N_c^{0}}+ \bm{N_c^{-2}})\bm{C^{2}}\\
(\bm{N_c^{0}}+ \bm{N_c^{-2}})\bm{C^{1}} 
\end{bmatrix}
+
\begin{bmatrix}
\bm{N_c^{-1}}\bm{C^{K-1}}\\
\bm{N_c^{-1}}(\bm{C^{K}}+ \bm{C^{K-2}})\\
\bm{N_c^{-1}}(\bm{C^{K-1}}+ \bm{C^{K-3}})\\
\vdots\\
\bm{N_c^{-1}}(\bm{C^{3}}+ \bm{C^{1}})\\
\bm{N_c^{-1}}\bm{C^{2}}
\end{bmatrix}
\,.
\end{equation}
Taking the large-$N_c$ limit is equivalent to keeping only the leading order of $N_c$ in each row, and dropping terms going as $N_c^{-2}$ compared to the leading term. Translating this back to the form in Eq.~\eqref{matrix-representation}, this becomes
\begin{equation}\label{matrix-representation-Nc}
\frac{\dd}{\dd t}
\begin{bmatrix}
C^K\\
\bm{C^{K-1}}\\
\bm{C^{K-2}}\\
\vdots\\
\bm{C^{2}}\\
\bm{C^1}
\end{bmatrix}
\sim
\left(
\textrm{diag}\left(N_c \sigma,\dots,N_c \sigma \right)
+
\begin{bmatrix}
0 & \bm{0} & \bm0 & \dots  & \dots   & \bm 0\\
\bm{\sigma} & \bm{0} & \bm{0} & \bm{0} & \dots & \bm 0\\
\bm0 & \bm{\sigma} & \bm{0} & \bm{0} & \bm{0} & \bm 0\\
\vdots & \vdots & \vdots & \vdots & \vdots & \vdots \\
\bm{0} & \dots & \bm{0} & \bm{\sigma} & \bm{0} & \bm{0}\\
\bm{0} & \dots & \dots & \bm{0} & \bm{\sigma}&\bm0\\
\end{bmatrix}
\right)
\begin{bmatrix}
C^K\\
\bm{C^{K-1}}\\
\bm{C^{K-2}}\\
\vdots\\
\bm{C^{2}}\\
\bm{C^1}
\end{bmatrix}
\,.
\end{equation}
Hence, in the large-$N_c$ limit, all of the terms above the diagonal go to zero, and the system simplifies drastically. To get some more intuition into why this is true physically it is useful to imagine being in some color configuration that scales as $\sim N_c^M$. At any point it is possible to exchange one gluon, after which the possible resulting color configurations of the system will change its $N_c$ power by exactly one, and go as $\sim N_c^{M+1}$ or $\sim N_c^{M-1}$. The gluon exchange comes with a factor $\sigma\sim N_c^{-1}$, so in total the overall $N_c$ power of going to these systems are $N_c^M$ and $N_c^{M-2}$. In the large-$N_c$ approximation the latter possibility is discarded, which is equivalent to dropping all the terms above the diagonal in the matrix \eqref{matrix-representation-Nc}.

It is clear from this discussion that the system of differential equations \eqref{master-diff-eq} simplifies, in the large-$N_c$ limit, to 
\begin{align} \label{master-eq-largeNc}
\frac{\dd}{\dd t}C_{m_1m_2\dots m_i \dots m_j \dots m_K}^M \simeq 
&-\frac12 n(t) N_c\sum_{k=1}^K \sigma_{\bar k m_k}C_{m_1m_2\dots m_i \dots m_j \dots m_K}^M \nn
&-\frac12 n(t)\sum_{i=1}^{K-1}\sum_{j>i}^K (\sigma_{\bar i m_j}+\sigma_{m_i\bar j }-\sigma_{m_i m_j}-\sigma_{\bar i \bar j})C^{M+1}_{m_1m_2\dots m_j \dots m_i \dots m_K} \,.
\end{align}
Here we have included superscripts to show the $N_c$-scaling. In the second line, we have indicated that only the correlators scaling as $N_c^{M+1}$ should be included in the sum. This means that in the large-$N_c$ limit the correlators with $M$ traces only depend on the correlators with $M+1$ traces. Similarly, the correlators with $M+1$ traces depend on the correlators with $M+2$ traces and so on. This continues all the way up to the correlators with $K-1$ traces, which depend on the correlators with $K$ traces. Using \eqref{master-eq-largeNc} the differential equation for the correlator scaling as $N_c^K$ is 
\begin{align}\label{Q-N-traces}
\frac{\dd}{\dd t} C_{12\dots K}  &\simeq -\frac12 n(t) N_c \sum_{k=1}^K \sigma_{\bar k k}C_{12\dots K} \,.
\end{align}
Since this is exactly solvable,
\begin{equation}
C_{12\dots K}(t)=N_c^K \,\rme^{-\frac12 N_c \int^t_{t_2} \dd s \,n(s)\sum_{k=1}^K \sigma_{\bar k k}(s)}\,,
\end{equation}
this provides a ``bootstrap'' for the whole system of equations.
The above argument shows that in principle all the correlators can be solved exactly in the large-$N_c$ limit. 

As a side note, we can also understand the large-$N_c$ approximation as a simplification of the operation of performing medium averages on multiple traced correlators. Given that a dipole in the large-$N_c$ is given by
\begin{equation}
\Sc_{1\bar1}(t,t_2)\equiv \frac{1}{N_c}\langle \tr[V_1 V_{\bar 1}^\d] \rangle=\rme^{-\frac12 N_c \int_{t_2}^t \dd s \, n(s) \sigma_{1\bar1}} \,,
\end{equation}
the answer for $C_{1\ldots K}(t)$ is just given by the product of $K$ dipoles, i.e. $C_{1\dots K} \simeq N_c^K \Sc_{1\bar1}\dots \Sc_{K\bar K}$. On the level of the full correlator, this corresponds to the simplification
\begin{align} \label{Q-N-traces-dipoles}
\langle\tr[W^1]\tr[W^2]\dots\tr[W^K]\rangle \approx \langle\tr[W^1]\rangle\langle\tr[W^2]\rangle\dots\langle\tr[W^K]\rangle \,.
\end{align}
This argument can also be extended to any of the other correlators discussed above, e.g. $\langle \tr [W^1] \tr[W^2 \ldots W^K]\rangle \approx  \langle \tr[W^1] \rangle \langle \tr [W^2\ldots W^K]\rangle$.

The simplified differential equation, Eq.~\eqref{master-eq-largeNc}, can also be solved directly to get the recursive formula
\begin{align}
&C_{m_1m_2\dots m_i \dots m_j \dots m_K}^M = N_c^M\, \rme^{-\frac12 N_c \int_{t_2}^t \dd s \,n(s) \sum_{k=1}^K \sigma_{\bar k m_k}} \nn
&- \frac12 \int_{t_2}^t \dd s \, n(s) 
\sum_{i=1}^{K-1}\sum_{j>i}^K \left((\sigma_{\bar i m_j}+\sigma_{m_i\bar j }-\sigma_{m_i m_j}-\sigma_{\bar i \bar j})C^{M+1}_{m_1m_2\dots m_j \dots m_i \dots m_K}\right)\nn
&\times \rme^{-\frac12 N_c \int_{s}^t \dd s' \,n(s') \sum_{k=1}^K \sigma_{\bar k m_k}}\,.
\end{align}
This can also be written in terms of dipoles, namely
\begin{align}\label{largeNc-solved}
&C_{m_1m_2\dots m_i \dots m_j \dots m_K}^M = N_c^M \prod_{k=1}^K \Sc_{m_k \bar k}(t,t_2) \nn
&- \frac12 \int_{t_2}^t \dd s \, n(s) 
\sum_{i=1}^{K-1}\sum_{j>i}^K \left((\sigma_{\bar i m_j}+\sigma_{m_i\bar j }-\sigma_{m_i m_j}-\sigma_{\bar i \bar j})C^{M+1}_{m_1m_2\dots m_j \dots m_i \dots m_K}\right)
\prod_{k=1}^K \Sc_{m_k \bar k}(t,s)\,.
\end{align}
From this equation it is clear that all of the Wilson line correlators can be written in terms of dipoles in the large-$N_c$ limit. That is because Eq.~\eqref{largeNc-solved} is a recursive relation (``bootstrap'') that stops when you reach the term with $K$ traces, which is given in terms of dipoles in \eqref{Q-N-traces-dipoles}. Since the only Wilson line structure that appears in both \eqref{Q-N-traces-dipoles} and \eqref{largeNc-solved} is dipoles, it means all the correlator can be written in terms of dipoles. In Ref.~\cite{Dominguez:2012ad} it was pointed out that that all higher-order correlators can be reduced to dipoles and quadrupoles at large-$N_c$. The result in this section directly confirms this, and show that really only dipoles are needed.

We could, in principle, also devise a scheme to compute sub-leading color corrections, that scale like $N_c^{-2}$ relative to the leading terms, following the steps in Eqs.~\eqref{eq:largeNc-correction-1} and \eqref{eq:largeNc-correction-2}, and below. We have nevertheless not pursued this program further in this work.

\section{Conclusion and outlook}
\label{sec:conclusions}

In this paper we have developed a general method for calculating correlators involving an arbitrary number of Wilson lines in the fundamental representation. This culminated in the system of differential equations in Eq.~\eqref{master-diff-eq}. This system can be solved numerically. We showed that in the large-$N_c$ limit the resulting simplified system of differential equations, Eq.~\eqref{master-eq-largeNc}, can be solved exactly. We also provided a general way to compute color sub-leading corrections, suppressed by $N_c^{-2}$ relative to the leading terms. This was done in detail for the four-point correlator, in Eqs.~\eqref{eq:largeNc-correction-1} and \eqref{eq:largeNc-correction-2}, but can easily be extended to any higher-order correlator. All the results can then be written in terms of dipoles and their convolutions.

This technique was applied on three different cases of $1 \to 2$ parton splittings in the medium, which were shown to involve correlators containing up to eight (fundamental) Wilson lines. We used our method to calculate these both at finite and large $N_c$. Comparisons of the results are shown in Fig.~\ref{fig:lund-ratio}. From these plots it is clear that in this exact case the large-$N_c$ approximations works quite well for small $\theta$, but the differences become bigger as $\theta$ grows. In certain areas of the phase space the error in using the large-$N_c$ limit might be as high as $16\%$. This is expected given that the corrections we find generically scale as $N_c^{-2}$.

Since our method deals with a generic set of correlated Wilson lines, representing particles moving on eikonal trajectories through a background field, it could easily be extended to many other physical situations.
For future work it would be interesting to apply our results in initial state physics, where similar correlators of Wilson lines also appear, and for soft contributions to event or jet observables in electron-positron or proton-proton collisions. Finally, we plan on extending the formulation to account for non-eikonal corrections to the particle trajectories.

\acknowledgments{We would like to thank A.~Takacs for useful discussions. This work is supported by a Starting Grant from Trond Mohn Foundation (BFS2018REK01) and the University of Bergen. }

\appendix
\section{Calculation of spectrums}\label{appendix:calc-of-spectrums}
Here we will show the calculations leading up to the for the emission spectra $\frac{\dd I}{\dd z \dd \theta}$. The Feynman rules from \cite{Mehtar_Tani_2018} have been used to calculate the matrix elements.

\subsection{Pair production}
We start with the process of a photon producing a quark-antiquark pair. This process has been calculated in \cite{Dom_nguez_2020} but we will restate some of the results. The amplitude is 
\begin{align}
\Mc_{s_1,s_2}^{ij} &= \int_{\p_0,\p_1',\p_2'} \int_{t_0}^L \dd t_1 \, (2\pi)^2 \delta(\p_0-\p_1'-\p_2')[(\p_1|\Gc_F(L,t_1)|\p_1')(\p_2'|\bar{\Gc}_F(L,t_1)|\p_2)]^{ij} \nn
&\times A_{\lambda,s_1,s_2}(\p_2'-z \p_0,z) \frac{1}{2 E} \rme^{-i \frac{\p_0^2}{2 E}(t_1-t_0)} \Mc_{0 \lambda}(\p_0)\,,
\end{align}
where the photon-quark vertex is given by
\begin{equation}
A_{\lambda,s_1,s_2}(\q,z)=\frac{2i e}{\sqrt{z(1-z)}}\delta_{-s_2s_1}(z\delta_{\gamma s_1}-(1-z)\delta_{\gamma s_2}) \q \cdot \epsilon_\gamma \,.
\end{equation}
The initial hard process is represented by the amplitude $\Mc_0$. After using the eikonal approximation \eqref{eikonal-prop-mom} this becomes (up to some phase that cancels when we take the square)
\begin{align}
\Mc_{s_1,s_2}^{ij} &= \frac{1}{2 E}\int_{t_0}^L \dd t_1 \,e^{i \frac{1}{2 z(1-z)E}((1-z)\p_2-z\p_1)^2 t_1}[V_1(L,t_1)V_2^\d(t_1,L)]^{ij} \nn
&\times A_{\lambda,s_1,s_2}((1-z)\p_2-z \p_1,z) \Mc_{0 \lambda}(\p_1+\p_2)\,.
\end{align}
We have used the more compact notation to write $V_F(\r_1) \equiv V_{1}$, $V_F(\r_2) \equiv V_{2}$. The goal is to calculate
\begin{equation}
\frac{\dd I}{\dd z \,\dd \theta} = \frac{z(1-z)E^2 \theta}{8 \pi^2} \frac{\left\langle|\Mc|^2\right\rangle}{\left\langle|\Mc_0|^2\right\rangle}\,.
\end{equation}
The Wilson lines can be split using $V(L,t_1)=V(L,t_2)V(t_2,t_1)$. Then we only need to deal with the two time intervals $(L,t_2)$ and $(t_2,t_1)$. After squaring the amplitude, averaging over initial polarization, summing the final spins, flavor and colors and taking the medium average this becomes \eqref{eq:spectrum-generic-profile} with \eqref{eq:c4-photon-splitting} and \eqref{eq:c3-photon-splitting}.

\subsection{Quark-gluon splitting}
The amplitude was calculated in  \cite{Mehtar_Tani_2018} and is
\begin{align}
\Mc_{\lambda,s}^{a i} &= \int_{\p_0,\p_0',\k',\p'} \int_{t_0}^L \dd t_1 \, (2\pi)^2 \delta(\p_0'-\k'-\p')(\k|\Gc_A^{ab}(L,t_1)|\k') \nn
&\times [(\p|\Gc_F(L,t_1)|\p')A^b_{\lambda,s,s'}(\k'-z \p_0',z)\frac{1}{2E}(\p_0'|\Gc_F(t_1,t_0)|\p_0)]^{ij} \Mc_{0 s'}^j(\p_0)\,,
\end{align}
where the quark-gluon vertex is
\begin{equation}
\begin{aligned}
A^{a i j}_{\lambda,s,s^{\prime}}(\boldsymbol{q}, z) = -\frac{2 i g \mathbf{t}^{a}_{i j}}{z \sqrt{1-z}} \delta_{s^{\prime} s}\left[\delta_{\lambda s}+(1-z) \delta_{\lambda-s}\right] \boldsymbol{q} \cdot \boldsymbol{\epsilon}_{\lambda}^{*}\,.
\end{aligned}
\end{equation}
Again this simplifies in the eikonal limit \eqref{eikonal-prop-mom}
\begin{equation}
\Mc_{\lambda,s}^{a i} = \frac{1}{2 E}\int_{t_0}^L \dd t_1 \,e^{i \frac{1}{2 z(1-z)E}((1-z)\k-z\p)^2 t_1}U_2^{ab}(L,t_1)[V_1(L,t_1)A^b_{\lambda,s,s'}V_0(t_1,t_0)]^{ij}
\Mc_{0 s'}^j(\k+\p)\,.
\end{equation}
We have denoted the adjoint Wilson line as $V_A(\r_2) \equiv U_{2}$. Squaring the amplitude, summing/averaging over spins and colors and taking the medium average gives
\begin{align} \label{dN-quark-gluon-adjoint}
\frac{\mathrm{d} I}{\mathrm{d} z \,\mathrm{d} \theta}&=\frac{\alpha_s}{\pi} \frac{P_{g q}(z)}{\theta}\frac{2}{N_c^2-1} 2 \operatorname{Re} \int_{t_0}^{L} \frac{\mathrm{d} t_1}{t_{\mathrm{f}}} \int_{t_1}^{L} \frac{\mathrm{d} t_2}{t_{\mathrm{f}}} \mathrm{e}^{-i \frac{t_2-t_1}{t_{\mathrm{f}}}} \nn
&\times \langle\left[U^\dagger(t_2,L)U(L, t_1)\right]^{\bar b b}
\tr\left[V_0^\dagger(0, t_2) t^{\bar b}V_{\bar 1}^\dagger(t_2,L)V_1(L,t_1) t^b V_0(t_1,0)\right]\rangle\,,
\end{align}
where the relevant Altarelli-Parisi splitting function is
\begin{equation}
P_{g q}(z)=C_F\frac{1+(1-z)^2}{z}\,.
\end{equation}
To continue we transform the adjoint Wilson lines into fundamental ones using the identity \eqref{eq:adj-fund-Wlines}.
The resulting expression will contain many group generators $t^a$, and can be simplified by using the Fierz identity \eqref{fierz}. Finally, completely in the fundamental representation the Wilson line structure becomes
\begin{align} \label{quark-gluon-W-lines}
&\left\langle\left[U^\dagger(t_2,L)U(L, t_1)\right]^{\bar b b}
\tr\left[V_0^\dagger(0, t_2) t^{\bar b}V_{\bar 1}^\dagger(t_2,L)V_1(L,t_1) t^b V_0(t_1,0)\right]\right\rangle \nn
&=\frac12\big\langle\left([V_2^\dagger V_{\bar 2} V_{\bar 1}^\dagger V_1]_{kj}[V_{\bar2}^\dagger V_2]_{il}-\frac{1}{N_c}[V_{\bar 1}^\dagger V_1]_{ij}\delta_{kl}\right)_{(L,t_2)}\nn
&\times\left([V_1 V_2^\dagger]_{jk}[V_2 V_0^\dagger]_{li}-\frac{1}{N_c}[V_1V_0^\dagger]_{ji}\delta_{lk}\right)_{(t_2,t_1)}\big\rangle\,.
\end{align}
Conservation of color then makes it possible to connect $i,l$ and $j,k$ so when we include the proper normalization factor the whole expression turns into \eqref{eq:spectrum-generic-profile} with \eqref{eq:c4-quark-gluon} and \eqref{eq:c3-quark-gluon}.

\subsection{Gluon-gluon splitting}
The calculation of the emission spectrum for gluon-gluon splittings was done in \cite{Blaizot_2013}. For completeness we will also include the main results here. The matrix element of the process is
\begin{align}
&\Mc_{\lambda_1,\lambda_2}^{a_1 a_2} = \int_{\k_0,\k_0',\k_1',\k_2'} \int_{t_0}^L \dd t_1 \, (2\pi)^2 \delta(\k_0'-\k_1'-\k_2')\nn
&\times (\k_1|\Gc_A^{a_1b_1}(L,t_1)|\k_1') (\k_2|\Gc_A^{a_2b_2}(L,t_1)|\k_2')A^{b_0b_1b_2}_{\lambda_0,\lambda_1,\lambda_2}(\k_2'-z \k_0',z)\frac{1}{2E}(\k_0'|\Gc_A^{b_0 c}(t_1,t_0)|\k_0)\nn
&\times \Mc_{0 \lambda_0}^c(\k_0)\,,
\end{align}
where the gluon-gluon vertex is
\begin{equation}
\begin{aligned}
A^{b_0 b_1 b_2}_{\lambda_0,\lambda_1,\lambda_2}(\boldsymbol{q}, z) = -2i g (T^{b_0})^{b_1b_2}\left[\frac{1}{z}(\q \cdot \bm \epsilon_{\lambda_2}^*)\delta_{\lambda_0\lambda_1}+ \frac{1}{1-z}(\q \cdot \bm \epsilon_{\lambda_1}^*)\delta_{\lambda_0\lambda_2}-(\q \cdot \bm \epsilon_{\lambda_0})\delta_{\lambda_1\lambda_2}\right]\,.
\end{aligned}
\end{equation}
In the eikonal approximation \eqref{eikonal-prop-mom} the amplitude is
\begin{align}
\Mc_{\lambda_1,\lambda_2}^{a_1 a_2} &= \frac{1}{2 E}\int_{t_0}^L \dd t_1 \,e^{i \frac{1}{2 z(1-z)E}((1-z)\k_2-z\k_1)^2 t_1}U_1^{a_1b_1}(L,t_1)U_2^{a_2b_2}(L,t_1) A^{b_0b_1b_2}_{\lambda_0,\lambda_1,\lambda_2} U_0^{b_0 c}(t_1,t_0) \nn
&\times \Mc_{0 s'}^c(\k_1+\k_2)\,.
\end{align}
After taking the square of the amplitude, summing/averaging over spins and colors and taking the medium average this becomes
\begin{align} 
\frac{\mathrm{d} I}{\mathrm{d} z \,\mathrm{d} \theta}&=\frac{\alpha_s}{\pi} \frac{P_{g g}(z)}{\theta}\frac{2}{N_c(N_c^2-1)} 2 \operatorname{Re} \int_{t_0}^{L} \frac{\mathrm{d} t_1}{t_{\mathrm{f}}} \int_{t_1}^{L} \frac{\mathrm{d} t_2}{t_{\mathrm{f}}} \mathrm{e}^{-i \frac{t_2-t_1}{t_{\mathrm{f}}}}  \nn
&\times f^{b_0 b_1 b_2}f^{\bar b_0 \bar b_1 \bar b_2} \langle [U_1^{a_1 d_1} U_2^{a_2 d_2}U_{\bar 1}^{\d \bar b_1 a_1}U_{\bar 2}^{\d \bar b_2 a_2}]_{(L,t_2)}
[U_1^{d_1 b_1}U_2^{d_2b_2}U_0^{\d b_0 \bar b_0}]_{(t_2,t_1)}\rangle\,,
\end{align}
where the relevant Altarelli-Parisi splitting function is
\begin{equation}
P_{g g}(z)=N_c\left[z(1-z)+\frac{1-z}{z}+\frac{z}{1-z}\right]\,.
\end{equation}
Conservation of color lets us decouple the Wilson lines in the two time intervals $(L,t_2)$ and $(t_2,t_1)$
\begin{equation}
f^{b_0 b_1 b_2}U_1^{d_1 b_1}U_2^{d_2b_2}U_0^{\d b_0 \bar b_0}
=\frac{1}{N_c(N_c^2-1)}f^{d_1 d_2 \bar b_0}f^{d_1' d_2' \bar b_0'}f^{b_0 b_1 b_2}U_1^{d_1' b_1}U_2^{d_2'b_2}U_0^{\d b_0 \bar b_0'}\,.
\end{equation}
The part in the time interval $(t_2,t_1)$ can be calculated explicitly because of its simple color structure
\begin{align}\label{ggg-Wilson-lines-t2t1}
\frac{1}{N_c(N_c^2-1)}f^{d_1' d_2' \bar b_0'}f^{b_0 b_1 b_2}U_1^{d_1' b_1}U_2^{d_2'b_2}U_0^{\d b_0 \bar b_0'}
= \rme^{-\frac{N_c}{2}\int_{t_1}^{t_2}\dd t\,n(t) [\sigma_{01}+\sigma_{02}+\sigma_{12}]}\,.
\end{align}
What remains are the Wilson lines in time interval $(L,t_2)$ 
\begin{equation}
f^{d_1 d_2 \bar b_0}f^{\bar b_0 \bar b_1 \bar b_2}\langle [U_1^{a_1 d_1} U_2^{a_2 d_2}U_{\bar 1}^{\d \bar b_1 a_1}U_{\bar 2}^{\d \bar b_2 a_2}]\rangle_{(L,t_2)}\,.
\end{equation}
However, these are not that easy to calculate. The procedure for calculating Wilson line products detailed in Sec. \ref{sec:Wilson} only involve fundamental Wilson lines, so \eqref{eq:adj-fund-Wlines} is used to turn all the adjoint Wilson lines into fundamental ones. Then one can use the definition of the structure constants $[t^a,t^b]=if^{abc}t^c$
and the identity \eqref{fierz} to get rid of all the group generators. This was done in \cite{Blaizot_2013}, and we quote the result
\begin{align} \label{ggg-Wilson-lines}
&f^{d_1 d_2 \bar b_0}f^{\bar b_0 \bar b_1 \bar b_2}\langle [U_1^{a_1 d_1} U_2^{a_2 d_2}U_{\bar 1}^{\d \bar b_1 a_1}U_{\bar 2}^{\d \bar b_2 a_2}]\rangle_{(L,t_2)}\nn
&=\frac12 \langle \tr[V_1V_{\bar 1}^\dagger]\tr[V_2 V_{\bar 2}^\dagger V_{\bar 1}V_1^\dagger]\tr[V_{\bar 2}V_2^\dagger]-\tr[V_1 V_{\bar 1}^\dagger V_2V_{\bar2}^\dagger V_{\bar1} V_1^\dagger V_{\bar 2}V_2^\dagger] +\textrm{h.c.}\rangle_{(L,t_2)} \nn
&= \langle \tr[V_1V_{\bar 1}^\dagger]\tr[V_2 V_{\bar 2}^\dagger V_{\bar 1}V_1^\dagger]\tr[V_{\bar 2}V_2^\dagger]-\tr[V_1 V_{\bar 1}^\dagger V_2V_{\bar2}^\dagger V_{\bar1} V_1^\dagger V_{\bar 2}V_2^\dagger]\rangle_{(L,t_2)}\,.
\end{align}
The last step is true because the medium averaged products of Wilson lines are real. This means that in the gluon-gluon case we end up with medium averaged products of up to eight Wilson lines. Putting it all together we get the formula \eqref{eq:spectrum-generic-profile} with \eqref{eq:c4-gluon-gluon-fund} and \eqref{eq:c3-gluon-gluon-fund}.

\section{Six and eight Wilson lines}\label{sec:six-eight-lines}
\subsection{Six lines}
In Sec. \ref{sec:Wilson} we developed the tools to calculate the correlators of six and eight Wilson lines, which appeared in \eqref{eq:c4-quark-gluon} and \eqref{eq:c4-gluon-gluon-fund}. To start we will look at the case of six lines, which follows from \eqref{master-diff-eq} with $K=3$. The relevant expression is $\frac{\dd}{\dd t}\langle [V_1 V_{\bar 1}^\dagger]_{i_1j_1}[V_2 V_{\bar 2}^\dagger]_{i_2j_2}[V_3 V_{\bar 3}^\dagger]_{i_3j_3}\rangle$. If this is contracted with $\delta_{j_1 i_2}\delta_{j_2 i_1}\delta_{j_3 i_3}$ it becomes
\begin{equation}
\delta_{j_1 i_2}\delta_{j_2 i_1}\delta_{j_3 i_3}\frac{\dd}{\dd t}\big\langle [V_1 V_{\bar 1}^\dagger]_{i_1j_1}[V_2 V_{\bar 2}^\dagger]_{i_2j_2}[V_3 V_{\bar 3}^\dagger]_{i_3j_3}\big\rangle
= \big\langle\tr[V_1 V_{\bar 1}^\dagger V_2 V_{\bar 2}^\dagger]\tr[V_3 V_{\bar 3}^\dagger]\big\rangle,
\end{equation}
which is the structure encountered in \eqref{eq:c4-quark-gluon}. To get exactly the same as in that equation we need only change the labels $(1,\bar 1,2,\bar 2,3,\bar 3)\to(1,2,\bar 2,\bar 1,2,\bar 2)$, which also simplifies the system somewhat. The six different projections are gathered into a vector
\begin{align}
\bm C^\intercal &= \big( C_{1\bar 22},C_{\bar 212},C_{2\bar 21},C_{12\bar 2},C_{21\bar 2},C_{\bar 221} \big)\nn
&=\Big(\langle\tr [V_1 V_2^\d]\tr [V_{\bar 2} V_{\bar 1}^\d]\tr [V_2 V_{\bar 2}^\d]\rangle,\langle 
\tr[V_1 V_2^\d V_{\bar 2} V_{\bar 1}^\d]\tr [V_2 V_{\bar 2}^\d] \rangle, \langle 
\tr[V_1 V_{\bar 2}^\d]\tr [V_{\bar 2} V_{\bar 1}^\d] \rangle,\nn
&\quad \langle \tr [V_1 V_2^\d] \tr[V_{\bar 1}^\d V_2] \rangle, \langle 
\tr[V_1 V_{\bar 1}^\d] \rangle, \langle 
\tr[V_1 V_2^\d V_{\bar 2} V_{\bar 1}^\d V_2 V_{\bar 2}^\d]\rangle \Big)\,.
\end{align}
We can write the system of differential equations as
\begin{equation} \label{diff-sys-3}
\frac{\dd}{\dd t}\bm C = -\frac12 n(t) \underline{\bm{\mathbb{M}}} \bm C \,.
\end{equation}
One can get the elements of the $6\times 6$ matrix $\underline{\bm{\mathbb{M}}}$ from \eqref{a-diagonal} and \eqref{a-non-diagonal}. The 6 diagonal entries are simply
\begin{align} \label{a-diagonal-3}
\mathbb{M}_{m_1m_2m_3}^{m_1m_2m_3}=N_c(\sigma_{2 m_1}+\sigma_{\bar 1 m_2}+\sigma_{\bar 2 m_3})-\frac{1}{N_c}\sigma_{1 \bar 1}\,,
\end{align}
where $(m_1,m_2,m_3)$ now is some permutation of $(1,\bar 2,2)$.
Thee non-zero non-diagonal entries are given by
\begin{align}
\mathbb{M}_{m_1m_2m_3}^{m_2m_1m_3}&=\sigma_{2 m_2}+\sigma_{\bar 1 m_1}-\sigma_{m_2m_1}-\sigma_{2 \bar 1}\nn
\mathbb{M}_{m_1m_2m_3}^{m_3m_2m_1}&=\sigma_{2 m_3}+\sigma_{\bar 2 m_1}-\sigma_{m_3m_1}-\sigma_{2 \bar 2}\nn
\mathbb{M}_{m_1m_2m_3}^{m_1m_3m_2}&=\sigma_{\bar 1 m_3}+\sigma_{\bar 2 m_2}-\sigma_{m_3m_2}-\sigma_{\bar 1 \bar 2}\,.
\end{align}
This leads to six differential equations which can be solved numerically for the six functions in $\bm C$. Interestingly this $6\times6$ system is reducible into two $3\times3$ systems. The first of these systems leads to three differential equations that actually can be solved exactly:
\begin{align}
\frac{\dd}{\dd t}\langle\tr[V_1 V_{\bar 1}^\d]\rangle =
&-C_F n(t) \sigma_{1\bar1}\langle\tr[V_1 V_{\bar 1}^\d]\rangle \nn
\frac{\dd}{\dd t}\langle\tr[V_1 V_{\bar 2}^\d]\tr [V_{\bar 2} V_{\bar 1}^\d]\rangle =
&-\frac12 n(t) [N_c(\sigma_{\bar1\bar2}+\sigma_{1 \bar2})-\frac{1}{N_c}\sigma_{1\bar1}]\langle\tr[V_1 V_{\bar 2}^\d]\tr [V_{\bar 2} V_{\bar 1}^\d]\rangle \nn
&-\frac12 n(t)(\sigma_{1\bar1}-\sigma_{1\bar2}-\sigma_{\bar1\bar2})\langle\tr[V_1 V_{\bar 1}^\d]\rangle\nn
\frac{\dd}{\dd t}\langle\tr[V_1 V_{2}^\d]\tr [V_{2} V_{\bar 1}^\d]\rangle =
&-\frac12 n(t) [N_c(\sigma_{\bar12}+\sigma_{1 2})-\frac{1}{N_c}\sigma_{1\bar1}]\langle\tr[V_1 V_{2}^\d]\tr [V_{2} V_{\bar 1}^\d]\rangle \nn
&-\frac12 n(t)(\sigma_{1\bar1}-\sigma_{12}-\sigma_{\bar12})\langle\tr[V_1 V_{\bar 1}^\d]\rangle
\end{align}
This is a nice consistency check, as taking the system for four Wilson lines \eqref{eq:diff-matrix-exact} and letting $2 \to \bar 2$ reproduces the first and second of these equations. Similarly, \eqref{eq:diff-matrix-exact} with $\bar 2 \to 2$ reproduces the first and third. Solving the first two gives
\begin{align}\label{3-Wilson-line-easy}
\langle\tr[V_1 V_{\bar 1}^\d]\rangle&=N_c\, \rme^{-C_F \int_{t_2}^t \dd s \,n(s) \sigma_{1\bar1}(s)}\nn
\langle\tr[V_1 V_{\bar 2}^\d]\tr [V_{\bar 2} V_{\bar 1}^\d]\rangle &=
(N_c^2-1)\,e^{-\frac12 \int_{t_2}^t \dd s \,n(s)(N_c(\sigma_{\bar1\bar2}+\sigma_{1\bar2})-\frac{1}{N_c}\sigma_{1\bar1})}+e^{-C_F \int_{t_2}^t \dd s \,n(s) \sigma_{1\bar1}(s)}\,.
\end{align}
One can easily get $\langle\tr[V_1 V_2^\d]\tr [V_2 V_{\bar 1}^\d]\rangle$ from the second of these equations by changing $\bar 2 \to 2$. The first equation in \eqref{3-Wilson-line-easy} is a well known result, so it is nice that we reproduce that.

\begin{figure}[H]
\begin{center}
\includegraphics[width=0.8\textwidth]{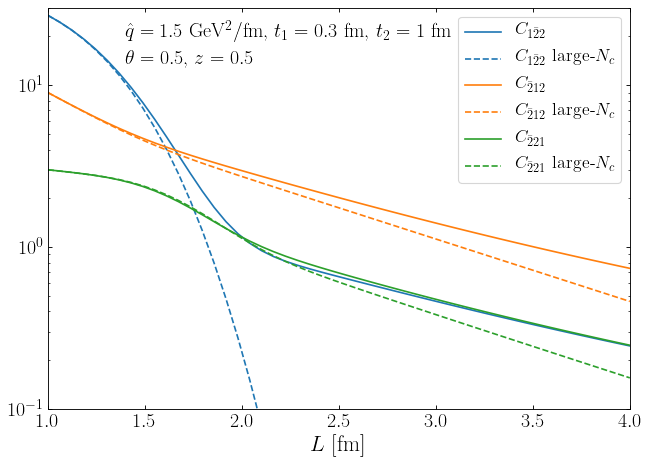} 
\end{center}
\caption{The exact and large-$N_c$ solutions to the system of differential equations \eqref{diff-sys-3-simplified}.}
\label{fig:3-Wilson-lines}
\end{figure}

However, this is nothing new, merely a check that the system of six Wilson lines is consistent with the previous calculations. The remaining $3\times3$ system contains the correlator we actually want to solve, but is also a more complicated nonhomogeneous system. It is useful to define two vectors with the 3 unknown and 3 known functions
\begin{align}
\bm C_1^\intercal &= \Big( \langle \tr [V_1 V_2^\d]\tr [V_{\bar 2} V_{\bar 1}^\d]\tr [V_2 V_{\bar 2}^\d] \rangle, \langle 
\tr[V_1 V_2^\d V_{\bar 2} V_{\bar 1}^\d]\tr [V_2 V_{\bar 2}^\d] \rangle, \langle 
\tr[V_1 V_2^\d V_{\bar 2} V_{\bar 1}^\d V_2 V_{\bar 2}^\d]\rangle \Big) \nn
\bm C_2^\intercal &= \Big(\langle \tr[V_1 V_{\bar 2}^\d]\tr [V_{\bar 2} V_{\bar 1}^\d] \rangle, \langle
\tr [V_1 V_2^\d] \tr[V_{\bar 1}^\d V_2] \rangle, \langle 
\tr[V_1 V_{\bar 1}^\d]\rangle \Big)\,.
\end{align}
Then we can write the remaining system of differential equations as
\begin{equation} \label{diff-sys-3-simplified}
\frac{\dd}{\dd t}\bm C_{1} = -\frac12 n(t) \left(\underline{\bm{\mathbb{M}}}_1 \bm C_1 +\underline{\bm{\mathbb{M}}}_2 \bm C_2\right)\,.
\end{equation}
The $3\times 3$ matrices $\underline{\bm{\mathbb{M}}}_1$ and $\underline{\bm{\mathbb{M}}}_2$ are subsets of the $6\times6$ matrix $\underline{\bm{\mathbb{M}}}$ and have the form 
\begin{equation}
\underline{\bm{\mathbb{M}}}_1=
\begin{bmatrix}
N_c(\sigma_{12}+\sigma_{\bar1\bar2}+\sigma_{2\bar2})-\frac{1}{N_c}\sigma_{1\bar1}& \sigma_{1\bar1}+\sigma_{2\bar2}-\sigma_{1\bar2}-\sigma_{\bar1 2}&0 \\
\sigma_{12}+\sigma_{\bar1\bar2}-\sigma_{1\bar2}-\sigma_{\bar1 2}& 2 (C_F \sigma_{1\bar1}+N_c\sigma_{2\bar2}) & \sigma_{1\bar2}+\sigma_{\bar1 2}-\sigma_{12}-\sigma_{\bar1\bar2}\\
0& \sigma_{1\bar1}+\sigma_{2\bar2}-\sigma_{12}-\sigma_{\bar1\bar2}&N_c(\sigma_{2\bar2}+\sigma_{1\bar2}+\sigma_{\bar1 2})-\frac{1}{N_c}\sigma_{1\bar1} 
\end{bmatrix}
\,.
\end{equation}

\begin{equation}
\underline{\bm{\mathbb{M}}}_2=
\begin{bmatrix}
\sigma_{1\bar 2}-\sigma_{12}-\sigma_{2\bar2}& \sigma_{\bar1 2}-\sigma_{2\bar2}-\sigma_{\bar1\bar2}& 0\\
0& 0& -2\sigma_{2\bar2}\\
\sigma_{\bar1\bar2}-\sigma_{2\bar2}-\sigma_{\bar1 2}& \sigma_{12}-\sigma_{1\bar2}-\sigma_{2\bar2}&0 
\end{bmatrix}
\,.
\end{equation}
This can be solved numerically for the three functions in $\bm C_1$, and the result can be seen in figure \ref{fig:3-Wilson-lines}.

\subsubsection{Quark-gluon splitting in the large-$N_c$}
As showed in section \ref{sec:Wilson-line-largeNc} all the functions in $\bm C$ can be solved exactly in the large-$N_c$ limit. The two terms with highest powers of $N_c$, $\langle[\tr [V_1 V_2^\d]\tr [V_{\bar 2} V_{\bar 1}^\d]\tr [V_2 V_{\bar 2}^\d]\rangle$ and $\langle\tr[V_1 V_2^\d V_{\bar 2} V_{\bar 1}^\d]\tr [V_2 V_{\bar 2}^\d]\rangle$ can be gotten directly from \eqref{Q-N-traces} and \eqref{largeNc-solved} respectively. Alternatively one can count the $N_c$ powers in \eqref{3-Wilson-line-easy} and realize that $\underline{\bm{\mathbb{M}}}_1$ and $\underline{\bm{\mathbb{M}}}_2$ simplify to
\begin{equation}
\underline{\bm{\mathbb{M}}}_1 \simeq
\begin{bmatrix}
N_c(\sigma_{12}+\sigma_{\bar1\bar2}+\sigma_{2\bar2})& 0 &0 \\
\sigma_{12}+\sigma_{\bar1\bar2}-\sigma_{1\bar2}-\sigma_{\bar1 2}& N_c (\sigma_{1\bar1}+2\sigma_{2\bar2}) &0\\
0& \sigma_{1\bar1}+\sigma_{2\bar2}-\sigma_{12}-\sigma_{\bar1\bar2}&N_c(\sigma_{2\bar2}+\sigma_{1\bar2}+\sigma_{\bar1 2}) 
\end{bmatrix}
\,.
\end{equation}

\begin{equation}
\underline{\bm{\mathbb{M}}}_2 \simeq
\begin{bmatrix}
0& 0& 0\\
0& 0& 0\\
\sigma_{\bar1\bar2}-\sigma_{2\bar2}-\sigma_{\bar1 2}& \sigma_{12}-\sigma_{1\bar2}-\sigma_{2\bar2}&0 
\end{bmatrix}
\,.
\end{equation}
The solutions to the simplified differential equation leads to \eqref{3-lines-largeNc}.

\subsection{Eight Wilson lines}
For more than six Wilson lines the matrix in \eqref{master-diff-eq} becomes so big that it is impractical to analyze it by hand. 

\begin{figure}[H]
\begin{center}
\includegraphics[width=0.8\textwidth]{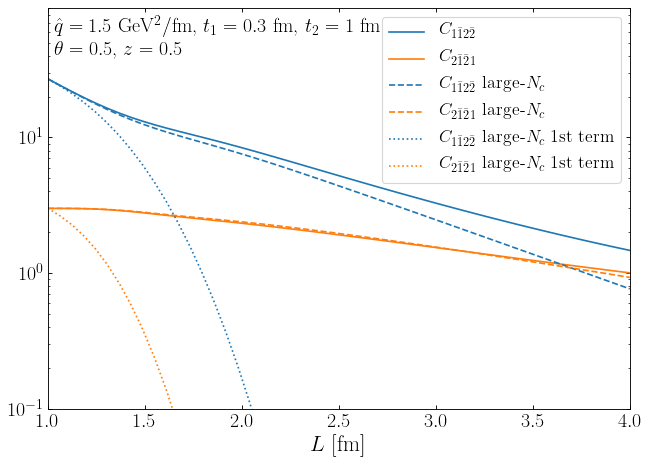} 
\end{center}
\caption{The exact and large-$N_c$ version of $C_{1\bar12\bar2}=\langle \tr[V_1V_{\bar 1}^\dagger]\tr[V_2 V_{\bar 2}^\dagger V_{\bar 1}V_1^\dagger]\tr[V_{\bar 2}V_2^\dagger]\rangle$ and $C_{2\bar1\bar2 2}=\langle\tr[V_1 V_{\bar 1}^\dagger V_2V_{\bar2}^\dagger V_{\bar1} V_1^\dagger V_{\bar 2}V_2^\dagger]\rangle$.}
\label{fig:8-Wilson-lines}
\end{figure}

For eight lines it involves the $4!=24$ projections of $\langle [V_1 V_{\bar 1}^\dagger]_{i_1j_1}[V_2 V_{\bar 2}^\dagger]_{i_2j_2}[V_3 V_{\bar 3}^\dagger]_{i_3j_3}[V_4 V_{\bar 4}^\dagger]_{i_4j_4}\rangle$, and the matrix $\underline{\bm{\mathbb{M}}}$ has $24^2$ elements. The power of our result in section \ref{sec:Wilson} is here evident, as simply solving the differential equation \eqref{master-diff-eq} for $K=4$ numerically immediately gives the result for eight lines Wilson lines. To get the Wilson line correlators we want from \eqref{eq:c4-gluon-gluon-fund} the four last labels must be changed $(3,\bar 3,4, \bar 4) \to (\bar 1,1,\bar 2,2)$ so that $\langle [V_1 V_{\bar 1}^\dagger]_{i_1j_1}[V_2 V_{\bar 2}^\dagger]_{i_2j_2}[V_3 V_{\bar 3}^\dagger]_{i_3j_3}[V_4 V_{\bar 4}^\dagger]_{i_4j_4}\rangle \to \langle [V_1 V_{\bar 1}^\dagger]_{i_1j_1}[V_2 V_{\bar 2}^\dagger]_{i_2j_2}[V_{\bar 1} V_1^\dagger]_{i_3j_3}[V_{\bar 2} V_2^\dagger]_{i_4j_4}\rangle$. The two relevant solutions are shown in figure \ref{fig:8-Wilson-lines}.

One thing to notice in figure \ref{fig:8-Wilson-lines} is that for the case of eight Wilson lines correlators, keeping only the first term in the large-$N_c$ limit does not work well.

\section{Derivation of differential equation} \label{appendix:proof-matrix-eq}
In this appendix we will show in more detail how the differential equation \eqref{master-diff-eq} was derived.

We start with the derivation of \eqref{K-Wilson-lines}. To illustrate we will first show the calculation for $K=2$, that is calculating $\langle[V_{1} V_{\bar 1}^\dagger]_{i_1j_1}[V_{2} V_{\bar 2}^\dagger]_{i_2j_2}\rangle$. This generalizes rather easily to the arbitrary $K$ case \eqref{K-lines}. Expanding the first of these pairs like in \eqref{Wilson-expansion} up to first order of $\epsilon$ and defining $\A \equiv A^a t^a$ it becomes
\begin{align}\label{2-Wilson-general}
&[V_1 V_{\bar 1}^\dagger]_{i_1 j_1}(t+\epsilon,t_0)=[V_1V^\dagger_{\bar 1} +ig\int^{t+\epsilon}_t \dd s(\A_1(s)V_1V^\dagger_{\bar 1}-V_1V^\dagger_{\bar 1}\A_{\bar 1}(s))\nn
&+\frac12 g^2 \int^{t+\epsilon}_t \dd s \int^{t+\epsilon}_t \rmd s' (2\A_1(s)V_1V^\dagger_{\bar 1}\A_{\bar 1}(s')-\A_1(s)\A_1(s')V_1V^\dagger_{\bar 1}-V_1V^\dagger_{\bar 1}\A_{\bar 1}(s)\A_{\bar 1}(s'))]_{i_1 j_1}\,.
\end{align}
Here all the Wilson lines on the right hand side go from $t_0$ to $t$. After taking the medium average \eqref{eq:med-avg} and using the Fierz identity \eqref{fierz} the last term becomes
\begin{align}
&\frac12 g^2 \int^{t+\epsilon}_t \dd s \int^{t+\epsilon}_t \rmd s'\langle [2\A_1(s)V_1V^\dagger_{\bar 1}\A_{\bar 1}(s')-\A_1(s)\A_1(s')V_1V^\dagger_{\bar 1}-V_1V^\dagger_{\bar 1}\A_{\bar 1}(s)\A_{\bar 1}(s')]_{i_1 j_1} \rangle \nn
&= \frac12 g^2 n(t)\epsilon\left[\gamma_{1\bar1}\langle\tr(V_1V_{\bar 1}^\d)\rangle \delta_{i_1 j_1}-(2 C_F \gamma_0 + \frac{1}{N_c} \gamma_{1\bar1})\langle V_1V_{\bar 1}^\d\rangle_{i_1 j_1}\right]\,.
\end{align}
Now adding the second pair of Wilson lines and taking the medium average, while disregarding higher orders of $\epsilon$, it takes the form
\begin{align}
&\langle[V_1 V_{\bar 1}^\dagger]_{i_1 j_1}[V_2 V_{\bar 2}^\dagger]_{i_2 j_2}\rangle(t+\epsilon,t_0)=\langle[V_1 V_{\bar 1}^\dagger]_{i_1 j_1}[V_2 V_{\bar 2}^\dagger]_{i_2 j_2}\rangle(t,t_0) \nn
&+\frac12 g^2 \epsilon \, n(t) \left\langle\left[\gamma_{1\bar1}\tr(V_1V_{\bar 1}^\d) \delta_{i_1 j_1}-(2 C_F \gamma_0 + \frac{1}{N_c} \gamma_{1\bar1})[V_1V_{\bar 1}^\d]_{i_1 j_1}\right][V_2V_{\bar 2}^\d]_{i_2 j_2}\right.\nn
&+[V_1V_{\bar 1}^\d]_{i_1 j_1}\left.\left[\gamma_{2\bar2}\tr(V_2V_{\bar 2}^\d) \delta_{i_2 j_2}-(2 C_F \gamma_0 + \frac{1}{N_c} \gamma_{2\bar2}) [V_2V_{\bar 2}^\d]_{i_2 j_2} \right]\right\rangle\nn
&-g^2\int^{t+\epsilon}_t \dd s \int^{t+\epsilon}_t \dd s' \big\langle\left[\A_1(s)V_1V_{\bar 1}^\d-V_1V_{\bar 1}^\d\A_{\bar 1}(s)\right]_{i_1 j_1}\left[\A_2(s')V_2V_{\bar 2}^\d-V_2V_{\bar 2}^\d\A_{\bar 2}(s')\right]_{i_2 j_2}\big\rangle\,.
\end{align}
The last term simplifies to
\begin{align}
&g^2\int^{t+\epsilon}_t \dd s \int^{t+\epsilon}_t \dd s'\big\langle\left[\A_1(s)V_1V_{\bar 1}^\d-V_1V_{\bar 1}^\d\A_{\bar 1}(s)\right]_{i_1 j_1}\left[\A_2(s')V_2V_{\bar 2}^\d-V_2V_{\bar 2}^\d\A_{\bar 2}(s')\right]_{i_2 j_2}\big\rangle \nn
&= \frac12 g^2 n(t) \epsilon \big\langle\gamma_{12}[V_1V_{\bar 1}^\d]_{i_2 j_1}[V_2V_{\bar 2}^\d]_{i_1 j_2}+\gamma_{\bar 1 \bar 2}[V_1V_{\bar 1}^\d]_{i_1 j_2}[V_2V_{\bar 2}^\d]_{i_2 j_1}\nn
&-\gamma_{1\bar2}\delta_{i_1 j_2}[V_2V_{\bar 2}^\d V_1V_{\bar 1}^\d]_{i_2 j_1}-\gamma_{\bar 1 2}\delta_{i_2 j_1}[V_1V_{\bar 1}^\d V_2V_{\bar 2}^\d]_{i_1 j_2}\nn
&+\frac{1}{N_c}(\gamma_{1\bar2}+\gamma_{\bar 1 2}-\gamma_{12}-\gamma_{\bar 1 \bar 2})[V_1V_{\bar 1}^\d]_{i_1 j_1}[V_2V_{\bar 2}^\d]_{i_2 j_2}\big\rangle\,.
\end{align}
Letting $\epsilon$ go to zero this turns into a differential equation
\begin{align} \label{4-lines-general}
&\frac{\dd}{\dd t}\langle[V_1 V_{\bar 1}^\dagger]_{i_1 j_1}[V_2 V_{\bar 2}^\dagger]_{i_2 j_2}\rangle \nn
&=\frac12 g^2 n(t)\big\langle[\frac{1}{N_c}(\gamma_{12}+\gamma_{\bar 1 \bar 2}-\gamma_{1\bar2}-\gamma_{\bar 1 2}-\gamma_{1\bar1}-\gamma_{2\bar2}-2(N_c^2-1)\gamma_0)][V_1 V_{\bar 1}^\dagger]_{i_1 j_1}[V_2 V_{\bar 2}^\dagger]_{i_2 j_2} \nn
&+\gamma_{1\bar1}\tr(V_1V_{\bar 1}^\d) \delta_{i_1 j_1}[V_2 V_{\bar 2}^\dagger]_{i_2 j_2}+\gamma_{2\bar2}\tr(V_2V_{\bar 2}^\d)[V_1 V_{\bar 1}^\dagger]_{i_1 j_1} \delta_{i_2 j_2}\nn
&-\gamma_{12}[V_1V_{\bar 1}^\d]_{i_2 j_1}[V_2V_{\bar 2}^\d]_{i_1 j_2}-\gamma_{\bar 1 \bar 2}[V_1V_{\bar 1}^\d]_{i_1 j_2}[V_2V_{\bar 2}^\d]_{i_2 j_1}\nn
&+\gamma_{1\bar2}\delta_{i_1 j_2}[V_2V_{\bar 2}^\d V_1V_{\bar 1}^\d]_{i_2 j_1}+\gamma_{\bar 1 2}\delta_{i_2 j_1}[V_1V_{\bar 1}^\d V_2V_{\bar 2}^\d]_{i_1 j_2}\big\rangle\,.
\end{align}
Now we only have to project out the two possible ways to connect the Wilson lines. Contracting with $\delta_{j_2i_1}\delta_{j_1i_2}$ and $\delta_{j_1 i_1}\delta_{j_2 i_2}$ gives $\frac{\dd}{\dd t}\langle\tr[V_1 V_{\bar 1}^\dagger V_2 V_{\bar 2}^\dagger]\rangle$ and $\frac{\dd}{\dd t}\langle\tr[V_1 V_{\bar 1}^\dagger]\tr[V_2 V_{\bar 2}^\dagger]\rangle$ respectively. In section \ref{sec:split-processes} we wanted to calculate $\langle\tr[V_1 V_2^\dagger V_{\bar 2} V_{\bar 1}^\dagger]\rangle$, which is similar to the above, but not exactly the same. Fortunately, our choice of labels is just a convention, and completely arbitrary. Simply making the three changes $\bar 1 \to 2$, $2 \to \bar 2$ and $\bar 2 \to \bar 1$ turns \eqref{4-lines-general} into the system of differential equations in \eqref{eq:diff-matrix-exact}. The difference in this approach compared to what we did in section \ref{sec:four-lines} is that \eqref{4-lines-general} contains both of \eqref{Q1-diff-eq} and \eqref{Q2-diff-eq}. This compact form is highly convenient when considering more than four Wilson lines. Generalizing the steps from equation \eqref{2-Wilson-general} to \eqref{4-lines-general} to an arbitrary number $K$ pairs of Wilson lines produces the differential equation \eqref{K-Wilson-lines}. 

The next step is to show how to get from Eq. \eqref{K-Wilson-lines} to the matrix elements \eqref{a-diagonal} and \eqref{a-non-diagonal}. Any pair of Wilson lines has two free indices. Take for example the second Wilson line pair in \eqref{K-Wilson-lines} which is $W_{i_2 j_2}^2$. Start by projecting out these two indices in all the ways possible, and at the same time making as few assumptions as possible about the rest of the Wilson lines. It turns out that projecting out with two Kronecker deltas gives all the information we need. There are also only two possibilities that need to be considered: either $W^2$ can connect to other Wilson lines, or it connects to itself and becomes a trace. These two possibilities are given by projecting with $\delta_{j_1 i_2}\delta_{j_2 i_3}$ and $\delta_{j_1 i_3}\delta_{j_2 i_2}$, respectively. To use Wilson lines 1, 2 and 3 is arbitrary. These labels can be changed to anything else without changing the result, so the calculation is completely general.

Using \eqref{K-Wilson-lines} and projecting out by the two deltas $\delta_{j_1 i_2}\delta_{j_2 i_3}$ gives a differential equation for $\langle[W^1 W^2 W^3]_{i_1 j_3}W_{i_4 j_4}^4\dots W_{i_K j_K}^K \rangle$.
\begin{align}
&\frac{\dd}{\dd t} \langle [W^1 W^2 W^3]_{i_1 j_3}W_{i_4 j_4}^4\dots W_{i_K j_K}^K \rangle   \nn
&=-\frac12 g^2 n(t)  \left(N_c(\sigma_{\bar 1 2}+\sigma_{\bar 2 3})+\frac{1}{N_c} A_K\right)\langle[W^1 W^2 W^3]_{i_1 j_3}W_{i_4 j_4}^4\dots W_{i_K j_K}^K \rangle \nn
&-\frac12 g^2 n(t)(\sigma_{\bar1 3}+\sigma_{2 \bar2}-\sigma_{23}-\sigma_{\bar1 \bar2})\langle\tr W^2 [W^1 W^3]_{i_1 j_3}W_{i_4 j_4}^4\dots W_{i_K j_K}^K \rangle\nn
&+ (\dots)\,.
\end{align}
The $(\dots)$ in the end are terms that are not completely determined by the projection that was made.

Next up is the case where we project out with $\delta_{j_1 i_3}\delta_{j_2 i_2}$, making a differential equation for $\langle\tr W^2 [W^1 W^3]_{i_1 j_3}W_{i_4 j_4}^4\dots W_{i_K j_K}^K \rangle$.
\begin{align}
&\frac{\dd}{\dd t} \langle\tr W^2 [W^1 W^3]_{i_1 j_3}W_{i_4 j_4}^4\dots W_{i_K j_K}^K \rangle \nn
&= -\frac12 g^2 n(t)  \left(N_c(\sigma_{\bar 1 3}+\sigma_{\bar 2 2})+\frac{1}{N_c} A_K\right)\langle\tr W^2 [W^1 W^3]_{i_1 j_3}W_{i_4 j_4}^4\dots W_{i_K j_K}^K \rangle \nn
&-\frac12 g^2 n(t) (\sigma_{\bar2 3}+\sigma_{2 \bar1}-\sigma_{23}-\sigma_{\bar1 \bar2}) \langle[W^1 W^2 W^3]_{i_1 j_3}W_{i_4 j_4}^4\dots W_{i_K j_K}^K \rangle\nn
&+ (\dots)\,.
\end{align}
In the notation from section \ref{sec:Wilson} these equations become
\begin{align}
\frac{\dd}{\dd t} C_{23 m_3 \dots m_N} &= -\frac12 g^2 n(t) \left(N_c(\sigma_{\bar 1 2}+\sigma_{\bar 2 3})+\frac{1}{N_c} A_K\right)C_{23 m_3 \dots m_N} \nn
&-\frac12 g^2 n(t) (\sigma_{\bar1 3}+\sigma_{2 \bar2}-\sigma_{23}-\sigma_{\bar1 \bar2}) C_{32 m_3 \dots m_N} \nn
&+ (\dots)\,,
\end{align}
\begin{align}
\frac{\dd}{\dd t} C_{32 m_3 \dots m_N} &
= -\frac12 g^2 n(t) \left(N_c(\sigma_{\bar 1 3}+\sigma_{\bar 2 2})+\frac{1}{N_c} A_K\right) C_{32 m_3 \dots m_N}  \nn
&-\frac12 g^2 n(t) (\sigma_{\bar2 3}+\sigma_{2 \bar1}-\sigma_{23}-\sigma_{\bar1 \bar2}) C_{23 m_3 \dots m_N} \nn
&+ (\dots)\,.
\end{align}
Both of these equations are consistent with the matrix elements \eqref{a-diagonal} and \eqref{a-non-diagonal}. The point is that when all the indices are projected out all the Wilson lines will connect in one of these two ways. Either they will connect to other Wilson lines or they will only connect to themselves. And since we have shown that in either way the resulting expression is given by \eqref{a-diagonal} and \eqref{a-non-diagonal} it means that these two equations are correct for all the possible combinations.

\bibliography{references.bib} 

\begin{thebibliography}{10}

\bibitem{dEnterria:2009xfs}
D.~d'Enterria, ``{Jet quenching},'' {\em Landolt-Bornstein}, vol.~23, p.~471,
  2010.

\bibitem{Majumder:2010qh}
A.~Majumder and M.~Van~Leeuwen, ``{The Theory and Phenomenology of Perturbative
  QCD Based Jet Quenching},'' {\em Prog. Part. Nucl. Phys.}, vol.~66,
  pp.~41--92, 2011.

\bibitem{Mehtar-Tani:2013pia}
Y.~Mehtar-Tani, J.~G. Milhano, and K.~Tywoniuk, ``{Jet physics in heavy-ion
  collisions},'' {\em Int. J. Mod. Phys. A}, vol.~28, p.~1340013, 2013.

\bibitem{Adams:2005dq}
J.~Adams {\em et~al.}, ``{Experimental and theoretical challenges in the search
  for the quark gluon plasma: The STAR Collaboration's critical assessment of
  the evidence from RHIC collisions},'' {\em Nucl. Phys. A}, vol.~757,
  pp.~102--183, 2005.

\bibitem{Adcox:2004mh}
K.~Adcox {\em et~al.}, ``{Formation of dense partonic matter in relativistic
  nucleus-nucleus collisions at RHIC: Experimental evaluation by the PHENIX
  collaboration},'' {\em Nucl. Phys. A}, vol.~757, pp.~184--283, 2005.

\bibitem{Aamodt:2010jd}
K.~Aamodt {\em et~al.}, ``{Suppression of Charged Particle Production at Large
  Transverse Momentum in Central Pb-Pb Collisions at $\sqrt{s_{NN}} =$ 2.76
  TeV},'' {\em Phys. Lett. B}, vol.~696, pp.~30--39, 2011.

\bibitem{Khachatryan:2016odn}
V.~Khachatryan {\em et~al.}, ``{Charged-particle nuclear modification factors
  in PbPb and pPb collisions at $ \sqrt{s_{\mathrm{N}\;\mathrm{N}}}=5.02 $
  TeV},'' {\em JHEP}, vol.~04, p.~039, 2017.

\bibitem{Chatrchyan:2011sx}
S.~Chatrchyan {\em et~al.}, ``{Observation and studies of jet quenching in PbPb
  collisions at nucleon-nucleon center-of-mass energy = 2.76 TeV},'' {\em Phys.
  Rev. C}, vol.~84, p.~024906, 2011.

\bibitem{Aad:2010bu}
G.~Aad {\em et~al.}, ``{Observation of a Centrality-Dependent Dijet Asymmetry
  in Lead-Lead Collisions at $\sqrt{s_{NN}}=2.77$ TeV with the ATLAS Detector
  at the LHC},'' {\em Phys. Rev. Lett.}, vol.~105, p.~252303, 2010.

\bibitem{Abelev:2013kqa}
B.~Abelev {\em et~al.}, ``{Measurement of charged jet suppression in Pb-Pb
  collisions at $\sqrt{s_{NN}}$ = 2.76 TeV},'' {\em JHEP}, vol.~03, p.~013,
  2014.

\bibitem{Baier:1994bd}
R.~Baier, Y.~L. Dokshitzer, S.~Peigne, and D.~Schiff, ``{Induced gluon
  radiation in a QCD medium},'' {\em Phys. Lett. B}, vol.~345, pp.~277--286,
  1995.

\bibitem{Baier:1996sk}
R.~Baier, Y.~L. Dokshitzer, A.~H. Mueller, S.~Peigne, and D.~Schiff,
  ``{Radiative energy loss and p(T) broadening of high-energy partons in
  nuclei},'' {\em Nucl. Phys. B}, vol.~484, pp.~265--282, 1997.

\bibitem{Baier:1996kr}
R.~Baier, Y.~L. Dokshitzer, A.~H. Mueller, S.~Peigne, and D.~Schiff,
  ``{Radiative energy loss of high-energy quarks and gluons in a finite volume
  quark - gluon plasma},'' {\em Nucl. Phys. B}, vol.~483, pp.~291--320, 1997.

\bibitem{Baier:1998yf}
R.~Baier, Y.~L. Dokshitzer, A.~H. Mueller, and D.~Schiff, ``{Radiative energy
  loss of high-energy partons traversing an expanding QCD plasma},'' {\em Phys.
  Rev. C}, vol.~58, pp.~1706--1713, 1998.

\bibitem{Zakharov:1996fv}
B.~G. Zakharov, ``{Fully quantum treatment of the Landau-Pomeranchuk-Migdal
  effect in QED and QCD},'' {\em JETP Lett.}, vol.~63, pp.~952--957, 1996.

\bibitem{Zakharov:1997uu}
B.~G. Zakharov, ``{Radiative energy loss of high-energy quarks in finite size
  nuclear matter and quark - gluon plasma},'' {\em JETP Lett.}, vol.~65,
  pp.~615--620, 1997.

\bibitem{Wiedemann:2000za}
U.~A. Wiedemann, ``{Gluon radiation off hard quarks in a nuclear environment:
  Opacity expansion},'' {\em Nucl. Phys. B}, vol.~588, pp.~303--344, 2000.

\bibitem{Mehtar_Tani_2018}
Y.~Mehtar-Tani and K.~Tywoniuk, ``Radiative energy loss of neighboring
  subjets,'' {\em Nuclear Physics A}, vol.~979, p.~165–203, Nov 2018.

\bibitem{Mehtar-Tani:2019tvy}
Y.~Mehtar-Tani, ``{Gluon bremsstrahlung in finite media beyond multiple soft
  scattering approximation},'' {\em JHEP}, vol.~07, p.~057, 2019.

\bibitem{Caucal:2018dla}
P.~Caucal, E.~Iancu, A.~H. Mueller, and G.~Soyez, ``{Vacuum-like jet
  fragmentation in a dense QCD medium},'' {\em Phys. Rev. Lett.}, vol.~120,
  p.~232001, 2018.

\bibitem{Caucal:2020xad}
P.~Caucal, E.~Iancu, A.~H. Mueller, and G.~Soyez, ``{Nuclear modification
  factors for jet fragmentation},'' {\em JHEP}, vol.~10, p.~204, 2020.

\bibitem{Dom_nguez_2020}
F.~Domínguez, J.~G. Milhano, C.~A. Salgado, K.~Tywoniuk, and V.~Vila,
  ``Mapping collinear in-medium parton splittings,'' {\em The European Physical
  Journal C}, vol.~80, Jan 2020.

\bibitem{Blaizot_2013}
J.-P. Blaizot, F.~Dominguez, E.~Iancu, and Y.~Mehtar-Tani, ``Medium-induced
  gluon branching,'' {\em Journal of High Energy Physics}, vol.~2013, Jan 2013.

\bibitem{Apolinario:2014csa}
L.~Apolin\'ario, N.~Armesto, J.~G. Milhano, and C.~A. Salgado,
  ``{Medium-induced gluon radiation and colour decoherence beyond the soft
  approximation},'' {\em JHEP}, vol.~02, p.~119, 2015.

\bibitem{Kovner_2001}
A.~Kovner and U.~A. Wiedemann, ``Eikonal evolution and gluon radiation,'' {\em
  Physical Review D}, vol.~64, Oct 2001.

\bibitem{Dominguez:2011wm}
F.~Dominguez, C.~Marquet, B.-W. Xiao, and F.~Yuan, ``{Universality of
  Unintegrated Gluon Distributions at small x},'' {\em Phys. Rev. D}, vol.~83,
  p.~105005, 2011.

\bibitem{Arnold:2019qqc}
P.~Arnold, ``{Landau-Pomeranchuk-Migdal effect in sequential bremsstrahlung:
  From large-$N$ QCD to $N$=3 via the SU($N$) analog of Wigner 6-$j$
  symbols},'' {\em Phys. Rev. D}, vol.~100, no.~3, p.~034030, 2019.

\bibitem{Zakharov:2018hfz}
B.~G. Zakharov, ``{Color randomization of fast gluon-gluon pairs in the
  quark-gluon plasma},'' {\em J. Exp. Theor. Phys.}, vol.~128, no.~2,
  pp.~243--258, 2019.

\bibitem{Hatta:2020wre}
Y.~Hatta and T.~Ueda, ``{Non-global logarithms in hadron collisions at $N_c$ =
  3},'' {\em Nucl. Phys. B}, vol.~962, p.~115273, 2021.

\bibitem{Jalilian-Marian:2004vhw}
J.~Jalilian-Marian and Y.~V. Kovchegov, ``{Inclusive two-gluon and valence
  quark-gluon production in DIS and pA},'' {\em Phys. Rev. D}, vol.~70,
  p.~114017, 2004.
\newblock [Erratum: Phys.Rev.D 71, 079901 (2005)].

\bibitem{Iancu:2011ns}
E.~Iancu and D.~N. Triantafyllopoulos, ``{Higher-point correlations from the
  JIMWLK evolution},'' {\em JHEP}, vol.~11, p.~105, 2011.

\bibitem{Lappi:2020srm}
T.~Lappi, H.~M\"antysaari, and A.~Ramnath, ``{Next-to-leading order
  Balitsky-Kovchegov equation beyond large $N_c$},'' {\em Phys. Rev. D},
  vol.~102, no.~7, p.~074027, 2020.

\bibitem{Dominguez:2012ad}
F.~Dominguez, C.~Marquet, A.~M. Stasto, and B.-W. Xiao, ``{Universality of
  multiparticle production in QCD at high energies},'' {\em Phys. Rev. D},
  vol.~87, p.~034007, 2013.

\bibitem{Nagy:2012bt}
Z.~Nagy and D.~E. Soper, ``{Parton shower evolution with subleading color},''
  {\em JHEP}, vol.~06, p.~044, 2012.

\bibitem{Hamilton:2020rcu}
K.~Hamilton, R.~Medves, G.~P. Salam, L.~Scyboz, and G.~Soyez, ``{Colour and
  logarithmic accuracy in final-state parton showers},'' 11 2020.

\bibitem{Gyulassy:2000er}
M.~Gyulassy, P.~Levai, and I.~Vitev, ``{Reaction operator approach to
  nonAbelian energy loss},'' {\em Nucl. Phys. B}, vol.~594, pp.~371--419, 2001.

\bibitem{Wang:1991xy}
X.-N. Wang and M.~Gyulassy, ``{Gluon shadowing and jet quenching in A + A
  collisions at s**(1/2) = 200-GeV},'' {\em Phys. Rev. Lett.}, vol.~68,
  pp.~1480--1483, 1992.

\bibitem{Aurenche:2002pd}
P.~Aurenche, F.~Gelis, and H.~Zaraket, ``{A Simple sum rule for the thermal
  gluon spectral function and applications},'' {\em JHEP}, vol.~05, p.~043,
  2002.

\bibitem{Mehtar_Tani_2020}
Y.~Mehtar-Tani and K.~Tywoniuk, ``Improved opacity expansion for medium-induced
  parton splitting,'' {\em Journal of High Energy Physics}, vol.~2020, Jun
  2020.

\bibitem{Barata:2020rdn}
J.~a. Barata, Y.~Mehtar-Tani, A.~Soto-Ontoso, and K.~Tywoniuk, ``{Revisiting
  transverse momentum broadening in dense QCD media},'' 9 2020.

\bibitem{Altinoluk_2014}
T.~Altinoluk, N.~Armesto, G.~Beuf, M.~Martınez, and C.~A. Salgado,
  ``Next-to-eikonal corrections in the cgc: gluon production and spin
  asymmetries in pa collisions,'' {\em Journal of High Energy Physics},
  vol.~2014, Jul 2014.

\end{thebibliography}
\bibliographystyle{ieeetr} 

\end{document}